\documentclass[a4paper,11pt]{article}
\pdfoutput=1 

\usepackage{jcappub} 

\usepackage[T1]{fontenc} 

\title{Testing Negative Effective Neutrino Mass Cosmologies with 21~cm Observations}

\author[a]{YiCheng Dai}
\author[a]{and Wei Liao}

\affiliation[a]{School of Physics, East China University of Science and Technology,\\
130 Meilong Road, Shanghai 200237, P. R. China}

\emailAdd{YichengDai@mail.ecust.edu.cn}
\emailAdd{liaow@ecust.edu.cn}

\abstract{
Recent baryon acoustic oscillation (BAO) measurements, when combined with cosmic microwave background (CMB) observations, have shown a preference for negative values of neutrino masses within the $\Lambda$CDM framework.
In this work, we investigate whether future 21 cm observations can distinguish cosmologies with negative effective neutrino masses favored by CMB and BAO data from a reference cosmology with $\sum m_\nu \simeq 0.06~{\rm eV}$.
Using two phenomenological extrapolation schemes, we obtain three representative cosmologies with negative effective neutrino mass 
whose predictions for CMB and BAO observables are consistent.
We then calculate the corresponding 21~cm power spectra and perform a Fisher matrix analysis for these cosmologies.
We find that 21~cm observations alone have limited ability to distinguish the three representative cosmologies from the reference cosmology, primarily due to degeneracies between neutrino mass and other  parameters.
Extending 21~cm observations to higher redshifts and imposing external Gaussian priors on cosmological parameters from 
CMB~+~BAO observations can reduce parameter degeneracies and improve the ability to distinguish cosmologies with negative neutrino mass from those with positive neutrino masses.
Our results highlight the importance of combining future 21 cm observations with complementary cosmological constraints from future CMB~+~BAO observations in testing the negative effective neutrino-mass preference.
}

\begin{document}
\maketitle
\flushbottom
\section{Introduction}
Neutrino oscillation experiments have established that neutrinos are massive and have measured the mass-squared differences between neutrino mass eigenstates. 
In particular, solar neutrino oscillation experiments,  atmospheric neutrino oscillation experiments and long baseline neutrino oscillation experiments have determined
$\Delta m_{21}^2\equiv m_2^2-m_1^2\simeq 7.5\times10^{-5}\,\mathrm{eV}^2$ and  $|\Delta m_{31}^2|\simeq 2.5\times10^{-3}\,\mathrm{eV}^2$.
For normal ordering ($m_1<m_2<m_3$), these measurements imply a lower bound on the sum of neutrino masses of $\sum m_\nu\gtrsim0.06\,\mathrm{eV}$, while for inverted ordering ($m_3<m_1<m_2$), the lower bound increases to $\sum m_\nu\gtrsim0.10\,\mathrm{eV}$~\cite{Esteban_2017, Gonzalez_Garcia_2021}.

Cosmological observations provide an independent probe of the absolute scale of neutrino masses.
The combination of cosmic microwave background (CMB) observations and baryon acoustic oscillation (BAO) measurements places stringent constraints on the sum of neutrino masses, $\sum m_\nu$.
An analysis combining BAO measurements from the Dark Energy Spectroscopic Instrument (DESI) Data Release 1 (DR1) with CMB data from the Planck and the Atacama Cosmology Telescope (ACT) obtained a stringent upper limit of $\sum m_\nu<0.072~\mathrm{eV}$ at 95\% confidence level (CL), within the $\Lambda$CDM framework and assuming three degenerate neutrino masses~\cite{Adame_2025}.
With the latest DESI DR2 BAO measurements, this limit has been further tightened to $\sum m_\nu<0.0642~\mathrm{eV}$ at 95\% CL when combined with CMB observations \cite{elbers2025constraintsneutrinophysicsdesi}.
The increasingly stringent cosmological upper limits on $\sum m_\nu$ suggest a growing tension with the lower bound $\sum m_\nu\gtrsim0.06\,\mathrm{eV}$ from neutrino oscillation experiments~\cite{Allali:2024aiv, Herold:2024nvk, Naredo-Tuero:2024sgf, Chebat:2025kes}.

Furthermore, some analyses of CMB and BAO data have shown a preference for negative values of $\sum m_\nu$~\cite{Chebat:2025kes, Craig:2024tky, Green:2024xbb, Elbers:2024sha}.
Since negative neutrino masses are not physically allowed, a phenomenological model of effective negative neutrino masses has been introduced to investigate this preference~\cite{Elbers:2024sha}.
 In this model, the physical neutrino energy is phenomenologically replaced by an effective energy,
\begin{equation}
E_{\rm eff}={\rm sgn}(m_{\nu})
\sqrt{p^2+m^2_\nu},
\label{eq: effective neutrino}
\end{equation}
where $p$ is the physical momentum of neutrino.
To preserve the standard radiation energy density, twice the corresponding positive relativistic energy density is added to the ultrarelativistic component to compensate for the negative effective neutrino energy density.
With this phenomenological model, we can investigate the effects of negative effective neutrino masses on the matter power spectrum.
For positive neutrino masses, neutrinos cannot cluster efficiently below a characteristic scale due to free streaming, leading to a suppression of the matter power spectrum. This scale is known as the free-streaming scale.
For negative effective neutrino masses, the free-streaming scale remains approximately unchanged for the same absolute neutrino mass and background expansion history.
However, unlike positive neutrino masses, negative effective neutrino masses lead to an enhancement of the small-scale matter power spectrum.
Some details are provided in Sec.~\ref{Effects of Negative Effective Neutrino Masses on Structure Formation and the 21 cm Signal}.

Since positive and negative effective neutrino masses have opposite effects on the small-scale matter power spectrum, they can consequently lead to different histories of structure formation.
These differences can affect the formation of early galaxies and the resulting 21 cm signal from neutral hydrogen in the intergalactic medium (IGM), during Cosmic Dawn and the Epoch of Reionization~\cite{Furlanetto:2006jb, Pritchard:2008wy, Shmueli:2023box}.
The potential of 21~cm cosmology to probe the high redshift Universe has motivated a growing number of observational efforts~\cite{Das:2018nw,Chen:2023hm,DeBoer:2016tnn,HERA:2025ajm,Mellema_2013}.
For instance, the Hydrogen Epoch of Reionization Array (HERA), designed to measure the 21~cm power spectrum during the Epoch of Reionization and Cosmic Dawn has been operating since 2017 and recently published its Phase~II results~\cite{DeBoer:2016tnn,HERA:2025ajm}.
Meanwhile, next-generation facilities, most notably the low-frequency array of the Square Kilometre Array (SKA-Low), are currently under construction and are expected to provide improved sensitivity to the high-redshift 21~cm power spectrum in the coming years~\cite{Mellema_2013}.
Therefore, future 21~cm observations are expected to help distinguish between cosmologies with positive and negative effective neutrino masses.

In this work, we first consider two phenomenological extrapolation schemes, Profile-Likelihood Extrapolation (PLE) and Observable-Level Extrapolation (OLE), to determine the cosmological parameters favored by CMB and BAO data in the region of negative effective neutrino mass.
We then calculate the corresponding 21 cm power spectra and investigate whether future 21~cm observations can distinguish these cosmologies from a reference cosmology with $\sum m_\nu\simeq0.06~\mathrm{eV}$ which is consistent with results from the neutrino oscillation experiments.
We find that 21 cm observations alone have limited ability to distinguish these cosmologies, mainly due to degeneracies between neutrino mass and other cosmological and astrophysical parameters.
However, introducing Gaussian priors on cosmological parameters based on other independent cosmological probes can reduce the impact of these degeneracies and improve the discriminating power, particularly between the OLE and reference cosmologies. 

This paper is organized as follows.
In Sec.~\ref{Cosmological Extrapolation Schemes}, we introduce the two phenomenological extrapolation schemes for negative effective neutrino masses and examine the consistency of their predictions for CMB and BAO observables.
In Sec.~\ref{21 cm Signal and Forecasts}, we first investigate the effects of negative effective neutrino masses on structure formation and the 21 cm signal.
We then perform Fisher matrix formalism for future 21 cm observations and investigate the impact of  external cosmological constraints.
Our findings are summarized in Sec.~\ref{Conclusions}.

\section{Cosmological Extrapolation Schemes}
\label{Cosmological Extrapolation Schemes}
Although the combination of CMB and BAO data prefers a negative value of the sum of neutrino masses, $\sum m_\nu$, such a value lies outside the physically allowed parameter space.
To phenomenologically describe the cosmology favored by CMB and BAO in the negative $\sum m_\nu$ region, we introduce two extrapolation schemes to obtain the corresponding cosmological parameters.

\subsection{Profile-Likelihood Extrapolation (PLE)}
The first scheme is the Profile-Likelihood Extrapolation (PLE).
We adopt the results for the benchmark case from our previous work~\cite{Dai:2026pko}, where the optical depth $\tau$ was treated as a free parameter.
For each fixed, non-negative value of $\sum m_\nu$, we used Procoli ~\cite{Karwal:2024qpt} to minimize $\chi^2$ over all other free parameters.
This gives the profiled curve $\chi^2_{\rm prof}(\sum m_\nu)$.
We then defined $\Delta\chi^2(\sum m_\nu)
=
\chi^2_{{\rm prof}}(\sum m_\nu)
-
\chi^2_{{\rm prof,min}}$, where $\chi^2_{{\rm prof,min}}$ is the minimum among the sampled profile points.
Based on this profile, we performed two quadratic fits using points with $\Delta\chi^2<4$ and $\Delta\chi^2<9$, respectively.
The minima of these two fits are located at $\sum m_\nu=-0.0196~{\rm eV}$ and $\sum m_\nu=-0.0368~{\rm eV}$, respectively.
The profile curves are shown in Fig.~2 of Ref.~\cite{Dai:2026pko}, and the corresponding extrapolation results are summarized in Tab.~1 of Ref.~\cite{Dai:2026pko}.

\begin{figure}[htbp]
\centering
\includegraphics[width=1\textwidth]{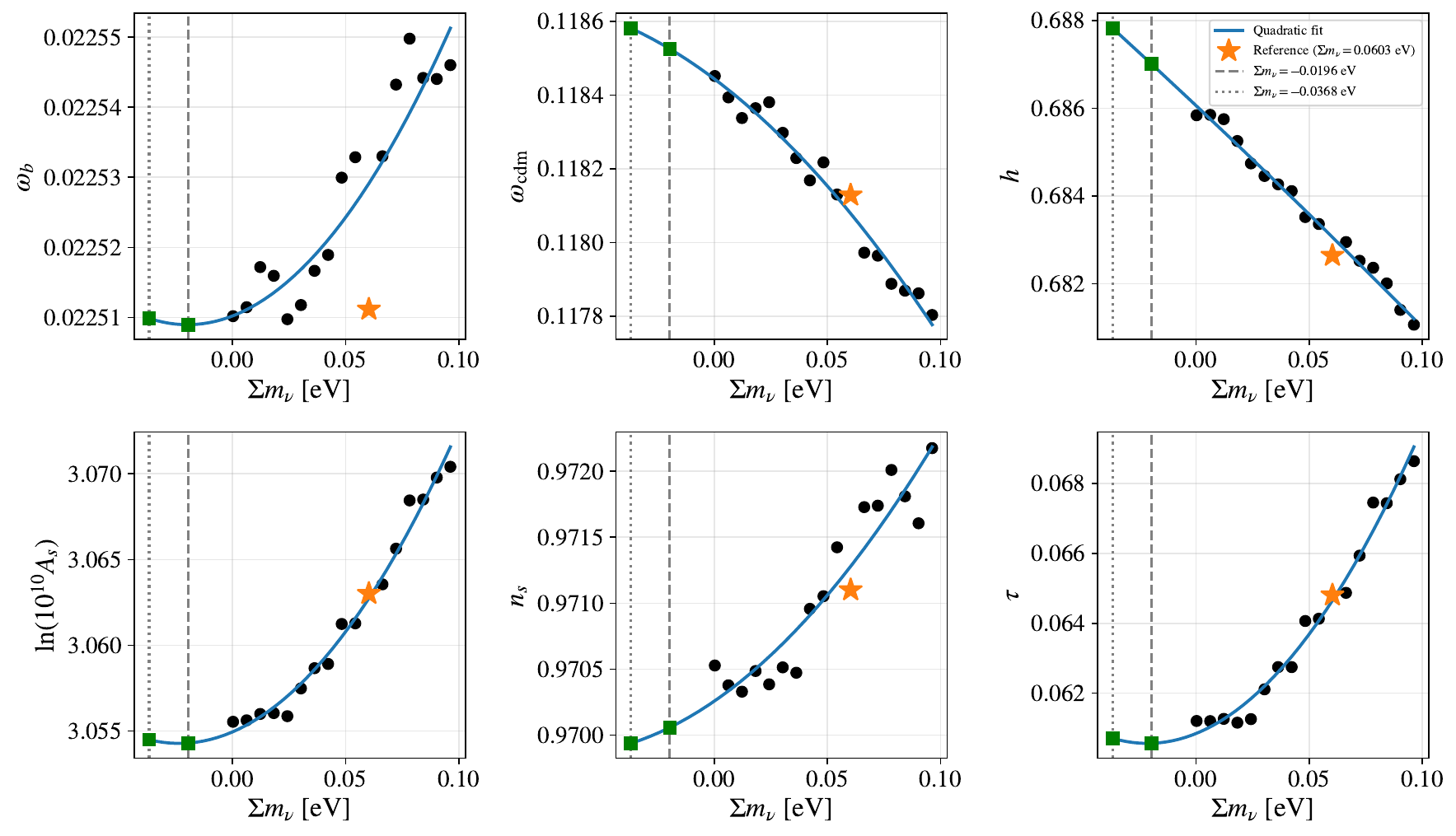}
    \caption{
Best-fit cosmological parameters as functions of $\sum m_\nu$.
Black dots show the best-fit values obtained at each fixed $\sum m_\nu$, while the orange stars highlight the reference cosmology with $\sum m_\nu=0.0603~{\rm eV}$.
The blue solid curve in each panel represents a quadratic fit to the black dots and orange star.
Green squares  in each panel indicate the parameter values obtained by extrapolating the quadratic fit to the two negative-mass minima $\sum m_\nu=-0.0196~{\rm eV}$ and $-0.0368~{\rm eV}$, marked by the vertical gray dashed and dotted lines, respectively.
}
    \label{fig: PLE_parameters}
\end{figure}

In addition to $\chi^2_{\rm prof}(\sum m_\nu)$, Procoli also yields the best-fit values of the remaining cosmological parameters at each fixed $\sum m_\nu$. We denote this parameter vector by $\boldsymbol{\theta}_{\rm prof}(\sum m_\nu)$, where
\begin{equation}
\boldsymbol{\theta}=
\{\omega_b,\omega_{\rm cdm},h,\tau,n_s,\ln(10^{10}A_s)\}.
\label{eq: theta}
\end{equation}
The best-fit parameter values at each fixed $\sum m_\nu$ are shown as black dots in Fig.~\ref{fig: PLE_parameters}.
As can be seen, they exhibit clear trends with neutrino mass.
This motivates us to extrapolate these parameters into the negative-mass region to estimate the corresponding cosmological parameters.
Specifically, we further fit each component of $\boldsymbol{\theta}_{\rm prof}(\sum m_\nu)$ with a quadratic function and extrapolate it to the two negative-mass minima, $\sum m_\nu=-0.0196~{\rm eV}$ and $\sum m_\nu=-0.0368~{\rm eV}$, respectively.
This gives two sets of extrapolated cosmological parameters.
The quadratic fit for each cosmological parameter is shown as a blue solid curve in the corresponding panel of Fig.~\ref{fig: PLE_parameters}, while the green squares indicate the values obtained by extrapolating the same fit to the two negative-mass minima.
For comparison with the negative effective neutrino-mass cosmologies, we adopt the sampled point at $\sum m_\nu=0.0603~{\rm eV}$, which is closest to $0.06~{\rm eV}$ as our reference cosmology.
Its cosmological parameters are listed in the last row of Tab.~\ref{tab: PLE_OLE_parameters}, and the corresponding point is highlighted by orange stars in Fig.~\ref{fig: PLE_parameters}.
We note that the orange star in the $\omega_b$ panel deviates noticeably from the quadratic fit. However, this deviation is less than 0.1\% in relative terms, about an order of magnitude smaller than the variations in the other cosmological parameters.
The specific values are listed in the second and third rows in Tab.~\ref{tab: PLE_OLE_parameters}.

These two parameter sets can be interpreted as phenomenological estimates of the cosmological parameters favored by CMB and BAO data under the PLE prescription.

\begin{table}[!htbp]
\centering
\renewcommand{\arraystretch}{1.15}
\setlength{\tabcolsep}{4pt}
\normalsize

\begin{tabular}{|l|l|c|c|c|c|c|c|c|}
\hline
Scheme
& $\sum m_\nu$ [eV]
& $\omega_b$
& $\omega_{\rm cdm}$
& $h$
& $\ln(10^{10}A_s)$
& $n_s$
& $\tau$
\\
\hline
PLE ($\Delta\chi^2<4$)
& $-0.0196$
& $0.0225$
& $0.1185$
& $0.6870$
& $3.0543$
& $0.9701$
& $0.0606$
\\
\hline
PLE ($\Delta\chi^2<9$)
& $-0.0368$
& $0.0225$
& $0.1186$
& $0.6878$
& $3.0545$
& $0.9699$
& $0.0607$
\\
\hline
OLE
& $-0.0793$
& $0.0225$
& $0.1190$
& $0.6892$
& $3.0438$
& $0.9682$
& $0.0560$
\\
\hline
Reference
& $0.0603$
& $0.0225$
& $0.1181$
& $0.6826$
& $3.0630$
& $0.9711$
& $0.0649$
\\
\hline
\end{tabular}

\caption{\label{tab: PLE_OLE_parameters}
Cosmological parameters at the best-fit points obtained from the two extrapolation schemes.
For PLE, the first and second rows correspond to parameter values extrapolated to $\sum m_\nu=-0.0196\,{\rm eV}$ and $-0.0368\,{\rm eV}$, respectively.
The procedure used to obtain these two mass values is described in the main text.
}
\end{table}

\subsection{Observable-Level Extrapolation (OLE)}
The second scheme is the Observable-Level Extrapolation (OLE), which adopts the extrapolation prescription proposed in Ref.~\cite{Elbers:2024sha}.
For a negative $\sum m_\nu$, we obtain the observables $X_{\boldsymbol{\theta}}(\sum m_\nu)$, such as the CMB power spectra and BAO distance observables, by linearly extrapolating the observables at zero neutrino mass and the corresponding positive mass $|\sum m_\nu|$,  while keeping the remaining cosmological parameters $\boldsymbol{\theta}$ fixed,
\begin{equation}
X_{\boldsymbol{\theta}}\left(\sum m_\nu\right)
=
X_{\boldsymbol{\theta}}\left(0\right)+{\rm sgn}\left(\sum m_\nu\right)\left[X_{\boldsymbol{\theta}}\left(\left|\sum m_\nu\right|\right)-X_{\boldsymbol{\theta}}\left(0\right)\right].
\label{eq: OLE_pre}
\end{equation}
For $\sum m_\nu>0$, this prescription recovers the ordinary physical neutrino model.
We implement this prescription through a CLASS wrapper, which applies Eq.~\eqref{eq: OLE_pre} to the numerical outputs returned by CLASS.

With this prescription, we perform cosmological fits to the combined CMB and BAO data using MontePython~\cite{Brinckmann:2018cvx, Audren:2012wb} and Procoli~\cite{Karwal:2024qpt}.
Unlike PLE, which extrapolates the best-fit cosmological parameters from the positive-mass region, OLE allows us to perform the fits directly in the negative-mass region.
The resulting global best-fit is located at $\sum m_\nu=-0.0793~{\rm eV}$, with the corresponding cosmological parameters listed in the fourth row of Tab.~\ref{tab: PLE_OLE_parameters}.

\subsection{Cosmological Fitting Setup}
\label{Cosmological Fitting Setup}
As discussed above, both extrapolation schemes involve cosmological fits.
In these fits, we use the same combination of CMB and BAO data.
For CMB data, we combine the Planck 2018 PR3~\cite{Planck2018results, Planck2018results_CMBpowerspectra} low-$\ell$ TT and EE likelihoods, the Planck 2018 Plik high-$\ell$ TTTEEE likelihood and the CMB lensing likelihoods from Planck PR4~\cite{Carron_2022}, ACT DR6~\cite{Qu_2024, Qu_2026}, and SPT-3G MUSE~\cite{Ge_2025}.
For BAO data, we use the DESI DR2 BAO distance measurements~\cite{elbers2025constraintsneutrinophysicsdesi, Abdul_Karim_2025}.

In both schemes, we vary the six standard cosmological parameters collected in $\boldsymbol{\theta}$, as defined in Eq.~\eqref{eq: theta}, together with the nuisance parameters associated with the observational likelihoods.
We assume three degenerate neutrino species, each with mass $m\equiv\frac{1}{3}\sum m_\nu$.

The theoretical observables are calculated using CLASS~\cite{DiegoBlas2011}, while cosmological fits are performed using MontePython~\cite{Brinckmann:2018cvx, Audren:2012wb} for Markov Chain Monte Carlo (MCMC) analyses and Procoli~\cite{Karwal:2024qpt} for profile likelihood analyses.

\subsection{Consistency of the Two Extrapolation Schemes}
\label{Consistency of the Two Extrapolation Schemes}
The second to the fourth rows of Tab.~\ref{tab: PLE_OLE_parameters} show the three CMB+BAO-preferred cosmologies obtained from the two extrapolation schemes.
Most of the parameters agree well among these extrapolation schemes.
Although some of the cosmological parameters differ noticeably, their predictions for the CMB and BAO observables are in good agreement.

For example, the values of $\tau$ in OLE is about $8\%$ lower than that in PLE.
Nevertheless, the combination $A_s\exp(-2\tau)$ remains nearly unchanged between PLE and OLE, with values of approximately $1.879\times10^{-9}$ and $1.876\times10^{-9}$, respectively.
Since this combination approximately characterizes the overall amplitude of the high-$\ell$ primary CMB power spectra, this agreement is noteworthy: the cosmological parameters at negative $\sum m_\nu$ in PLE are obtained through parameter extrapolation, whereas those in OLE are determined from a direct fit to the data in this region.

To further examine this agreement, we calculate the CDM~+~baryon power spectra $P_{cb}(k)$ in PLE scheme using a modified version of CLASS, which allows for negative neutrino masses.
Following the prescription in Ref.~\cite{Elbers:2024sha}, we replace the neutrino energy in the background and perturbation equations with Eq.~\eqref{eq: effective neutrino}.
To preserve the standard radiation energy density, we compensate for negative neutrino masses by adding twice the corresponding positive relativistic energy density to the ultrarelativistic component.
The density perturbation of neutrino with negative effective mass and its implications for the  CDM~+~baryon power spectrum are discussed 
in Sec.~\ref{Effects of Negative Effective Neutrino Masses on Structure Formation and the 21 cm Signal} and Appendix~\ref{Effective Neutrino Free Streaming}. 
We show that the free-streaming scale remains nearly unchanged when the sign of the neutrino mass is reversed.
However, negative effective neutrino masses enhance the small-scale matter power spectrum, rather than suppress it.

\begin{figure}[htbp]
\centering
\includegraphics[width=0.6\textwidth]{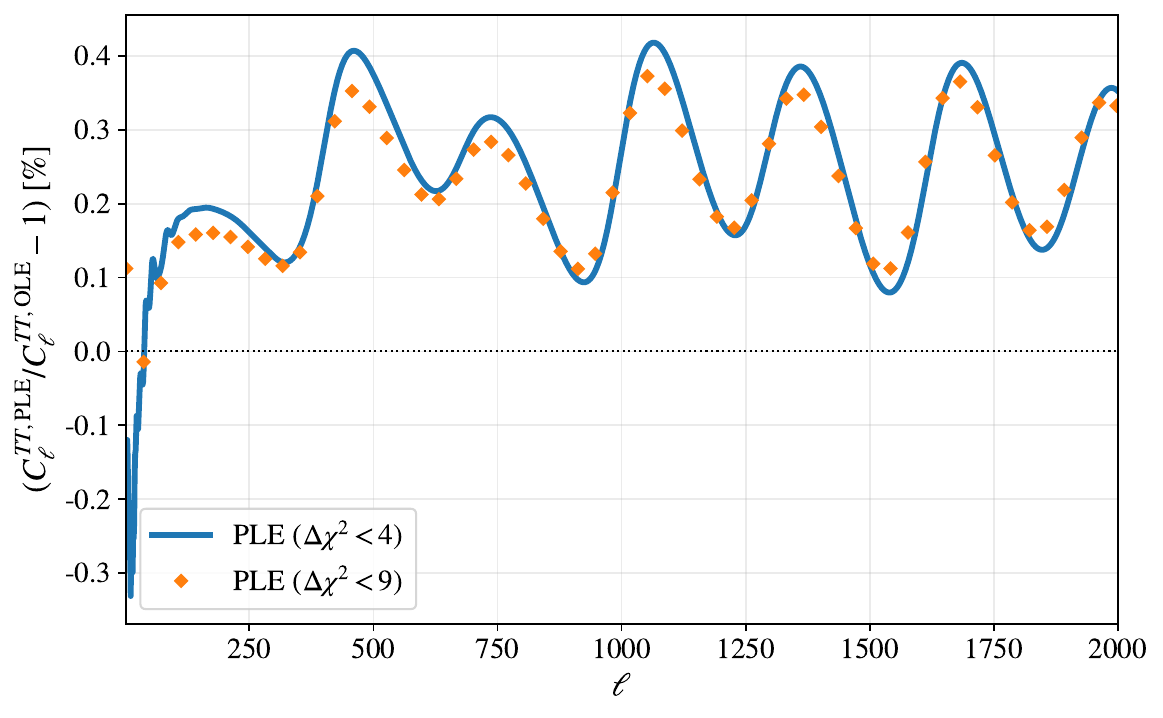}
    \caption{
Relative differences in the CMB TT angular power spectra between the two PLE cases and OLE, calculated using the cosmological parameters listed in Tab.~\ref{tab: PLE_OLE_parameters}.
The blue solid line represents the PLE $(\Delta\chi^2<4)$ case, while the orange diamonds represent the PLE $(\Delta\chi^2<9)$ case.
}
    \label{fig: TT}
\end{figure}

It should be noted that Ref.~\cite{Elbers:2024sha} has shown that the matter power spectrum calculated using the effective neutrino model, defined in Eq.~\eqref{eq: effective neutrino}, agrees well with that obtained from the observable-level extrapolation, defined in Eq~\eqref{eq: OLE_pre}.
In this subsection, we use the modified CLASS to calculate the CMB and BAO observables for the cosmologies obtained from PLE, while those for the OLE cosmology are obtained using Eq~\eqref{eq: OLE_pre}.
The resulting CMB TT angular power spectra in PLE are compared with those in OLE in Fig.~\ref{fig: TT}.
The relative differences between the two PLE cases and OLE remain below $0.5\%$ over the entire multipole range, indicating good agreement between the two extrapolation schemes.
This is consistent with the nearly identical values of $A_s\exp(-2\tau)$ discussed above.

We further compare the two schemes using the low-$\ell$ EE power spectrum.
The amplitude of the reionization bump is approximately characterized by the combination $A_s\tau^2$, which differs significantly between PLE and OLE, with values of approximately $7.80\times10^{-12}$ and $6.58\times10^{-12}$, respectively.
\begin{figure}[htbp]
\centering
\includegraphics[width=0.6\textwidth]{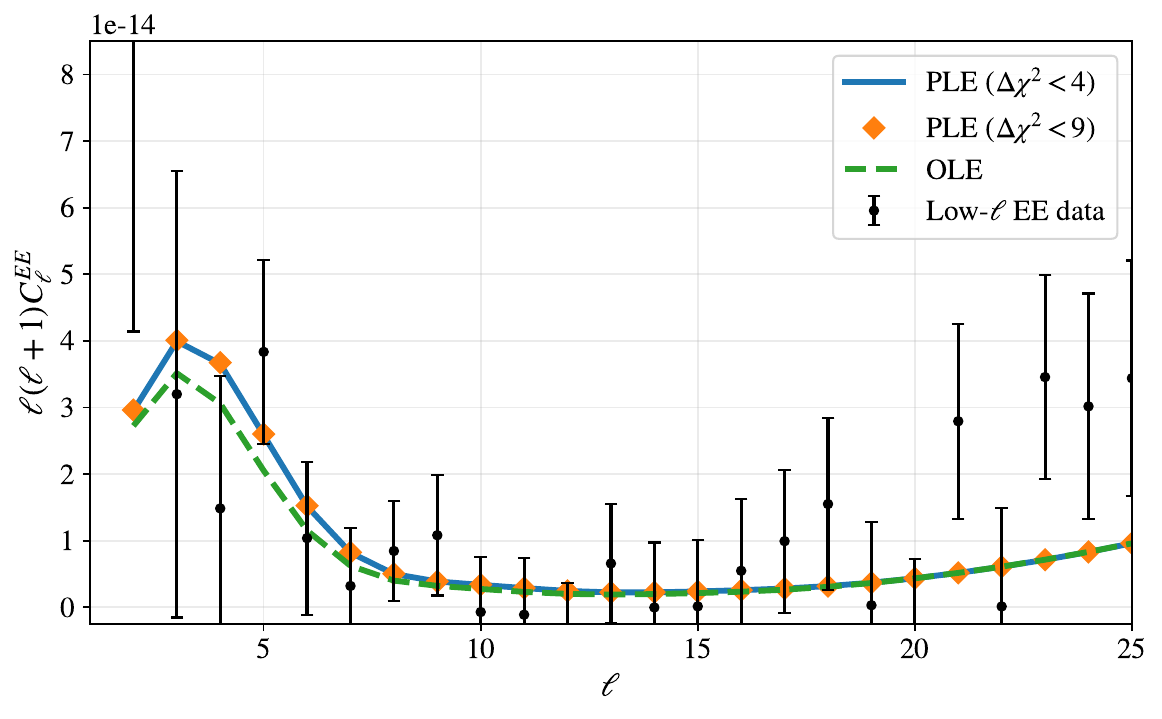}
    \caption{
The dimensionless low-$\ell$ EE spectra corresponding to the cosmological parameters listed in Tab.~\ref{tab: PLE_OLE_parameters}.
The blue solid line, the orange diamonds and the green dashed line represent the PLE $(\Delta\chi^2<4)$, PLE $(\Delta\chi^2<9)$ and OLE, respectively.
The black points and error bars are the data from Planck 2018~\cite{Planck2018results_CMBpowerspectra,Planck2018EEdata}.
}
    \label{fig: EE}
\end{figure}
The resulting low-$\ell$ EE power spectra are shown in Fig.~\ref{fig: EE}, together with the Planck 2018 data.
As shown by the PLE and OLE curves, the two schemes predict significant differences at $\ell\lesssim10$. 
However, the EE power spectrum is more than two orders of magnitude smaller than the TT spectrum, and the observational uncertainties remain large, as indicated by the error bars in Fig.~\ref{fig: EE}. 
The differences between the two schemes are mainly concentrated at multipoles where the error bars are large, and both the PLE and OLE predictions are consistent with the Planck data within these uncertainties.
Therefore, despite the difference in $A_s\tau^2$, the two schemes remain approximately consistent with the current low-$\ell$ EE observations.

The relative differences in the BAO distance observables are even smaller, remaining below $0.5\%$.
These results support the consistency of the two extrapolation schemes in their CMB and BAO predictions.
We thus obtain three representative cosmologies with negative effective neutrino masses favored by the CMB+BAO data.

\section{21 cm Signal and Forecasts}
\label{21 cm Signal and Forecasts}
The cosmologies with negative effective neutrino masses favored by CMB+BAO data may lead to different histories of structure formation compared with a cosmology with a physical positive neutrino mass due to neutrino free streaming,
and can affect the abundance of early halos and star formation, and consequently the 21 cm signal, as will be discussed in Sec.~\ref{Effects of Negative Effective Neutrino Masses on Structure Formation and the 21 cm Signal}.
In the remaining subsections, we perform a Fisher matrix analysis to investigate whether future 21 cm experiments can distinguish the three CMB+BAO-preferred cosmologies from a reference cosmology with a physical $\sum m_\nu=0.06~{\rm eV}$.

\subsection{Effects of Neutrino Mass on Structure Formation and the 21 cm Signal}
\label{Effects of Negative Effective Neutrino Masses on Structure Formation and the 21 cm Signal}
\begin{figure}[htbp]
\centering
\includegraphics[width=0.49\textwidth]{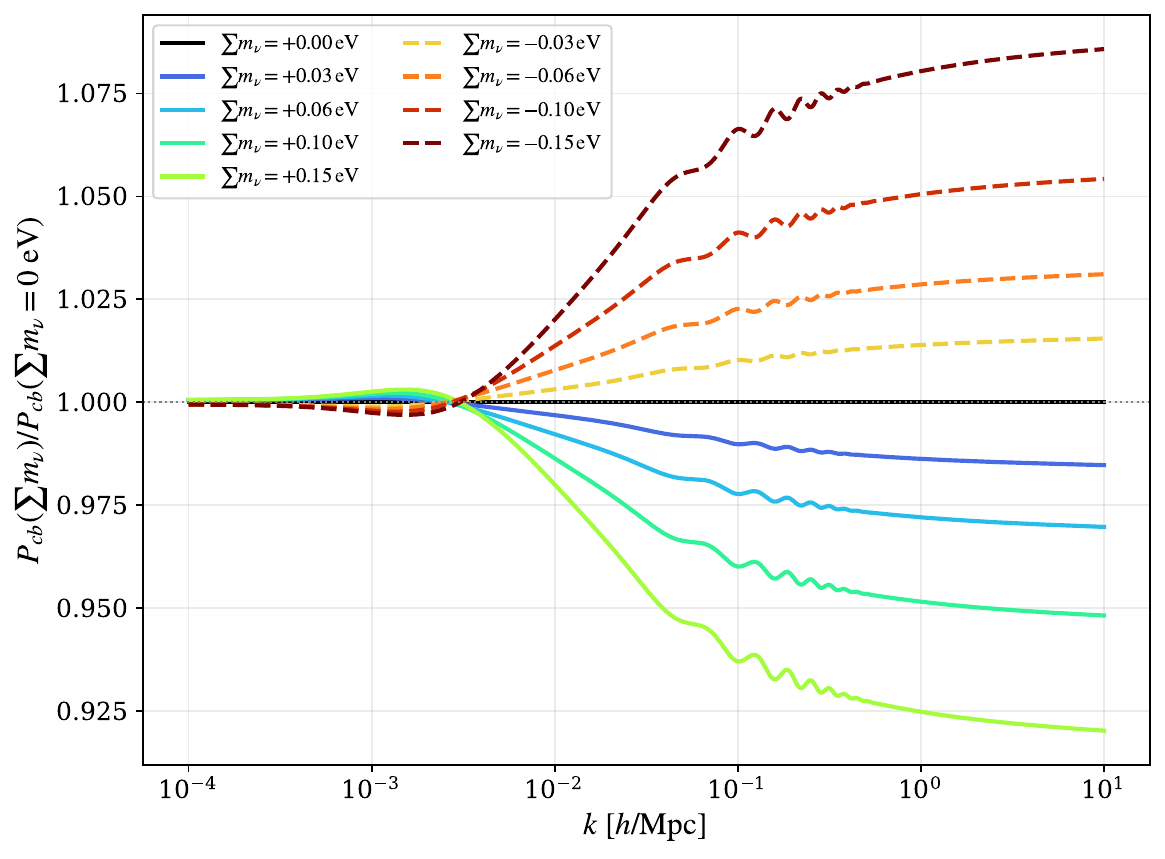}
    \caption{Ratio of the linear CDM~+~baryon power spectrum to that of the massless neutrino model at $z=0$.
Solid and dashed lines correspond to positive and negative effective neutrino masses, respectively.
}
    \label{fig: Pcb}
\end{figure}

To understand how negative effective neutrino masses affect the 21~cm signal, we first examine their impact on structure formation.
For clarity, in this subsection we assume fixed $\Omega_m$, $h$, and $\omega_b$.
The total matter density contrast is given by
\begin{equation}
\delta_m=f_c\delta_c+f_b\delta_b+f_\nu\delta_\nu,
\label{eq: delta_m_degenerate}
\end{equation}
where $f_i=\omega_i/\omega_m$ and $f_c+f_b+f_\nu=1$.
Here, $\delta_i\equiv\delta\rho_i/\bar{\rho}_i$ denotes the density contrast.
As shown by Eqs.~\eqref{eq: delta_nu_evol} and~\eqref{eq: freestreamingscale} in Appendix~\ref{Effective Neutrino Free Streaming}, neutrino density perturbations are suppressed by the pressure term on scales smaller than the free-streaming scale.
Consequently, the background density of the CDM~+~baryon component that clusters on these scales is modified by a factor $\bar{\rho}_c+\bar{\rho}_b=(1-f_\nu)\bar{\rho}_m$.
The evolution equation for the CDM+baryon density contrast then becomes
\begin{equation}
\ddot{\delta}_{cb}
+\frac{\dot a}{a}\dot{\delta}_{cb}
-4\pi G a^2(1-f_\nu)\bar{\rho}_m\delta_{cb}=0,
\label{eq: P_cb}
\end{equation}
where dots denote derivatives with respect to conformal time.
For positive neutrino masses, $f_\nu>0$, the gravitational source is reduced, 
leading to slower growth of CDM+baryon perturbations and a suppression of $P_{cb}(k)$.
In contrast, for negative effective neutrino masses, $f_\nu<0$, the gravitational source is enhanced, resulting in faster growth of CDM+baryon perturbations and thus an enhancement of $P_{cb}(k)$.
Fig.~\ref{fig: Pcb} shows the ratio of the CDM~+~baryon power spectrum to that of the massless neutrino model.
The solid lines show that positive neutrino masses suppress the CDM~+~baryon power spectrum  at small scales due to the free-streaming effect, while the dashed lines show that the same effect leads to an enhancement for negative effective neutrino masses.
The comparison between the solid and dashed lines confirms similar free-streaming wavenumbers $k_{\rm fs}$ for positive and negative effective neutrino masses with the same absolute value.

\begin{figure}[htbp]
\centering
\includegraphics[width=0.49\textwidth]{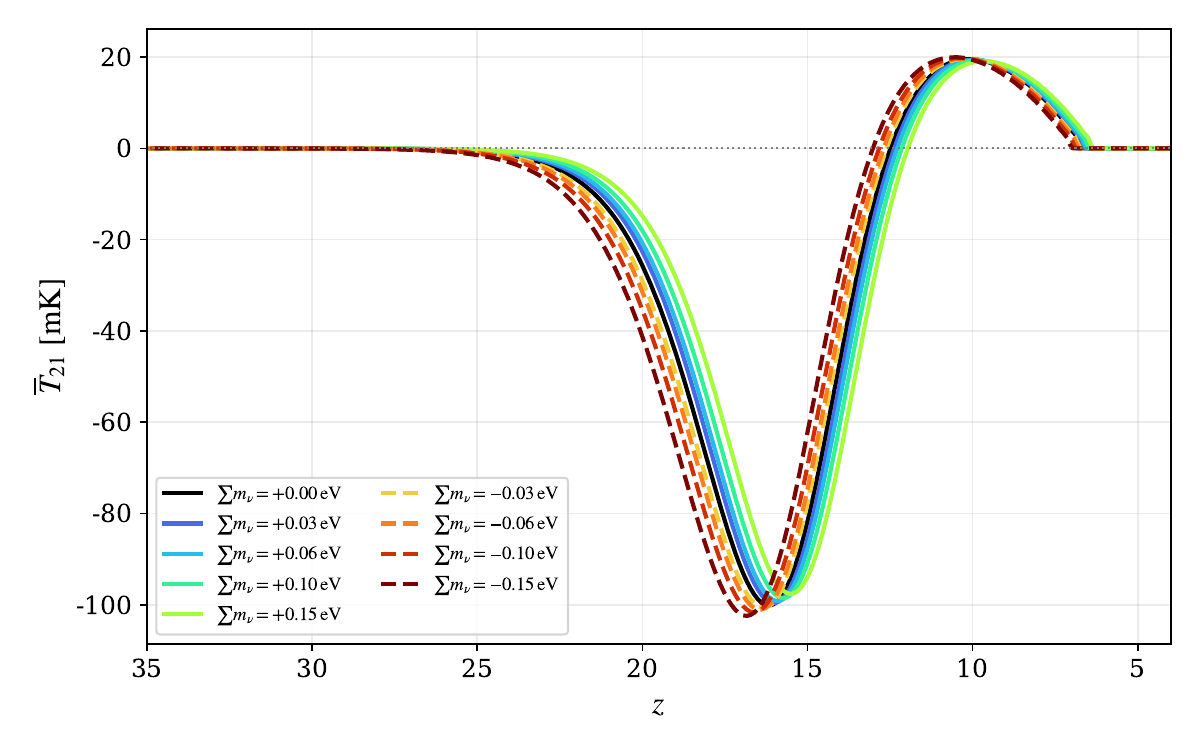}
    \caption{Global 21 cm signal $\bar{T}_{21}(z)$ as a function of redshift for different neutrino masses.
Solid and dashed lines correspond to positive and negative effective neutrino masses, respectively.
}
    \label{fig: 21cm_global_signal}
\end{figure}

The modification of the CDM~+~baryon power spectrum affects the abundance of dark matter halos, thereby influencing early galaxy formation. 
These changes also affect the thermal and ionization histories of the intergalactic medium (IGM), and thus the 21~cm signal.
The global 21 cm signal, $\bar{T}_{21}(z)$, denotes the mean brightness temperature of neutral hydrogen relative to the CMB background as a function of redshift.
The results of $\bar{T}_{21}(z)$ for different neutrino masses, 
with all other parameters held fixed,  are shown in Fig.~\ref{fig: 21cm_global_signal}.
The comparison between the solid and dashed curves indicates that, for a given absolute value of the effective neutrino mass, the features of the global 21~cm signal appear at slightly higher redshifts for negative effective masses.
This is because negative effective neutrino masses enhance the small scale $P_{cb}$, leading to earlier structure formation.
Therefore, for negative effective neutrino masses, the Wouthuysen-Field (WF) coupling becomes effective earlier, causing the global 21~cm signal to decrease at higher redshifts.
Meanwhile, X-ray heating also becomes dominant earlier so the absorption trough shifts to higher redshifts.

\begin{figure}[htbp]
\centering
\includegraphics[width=0.49\textwidth]{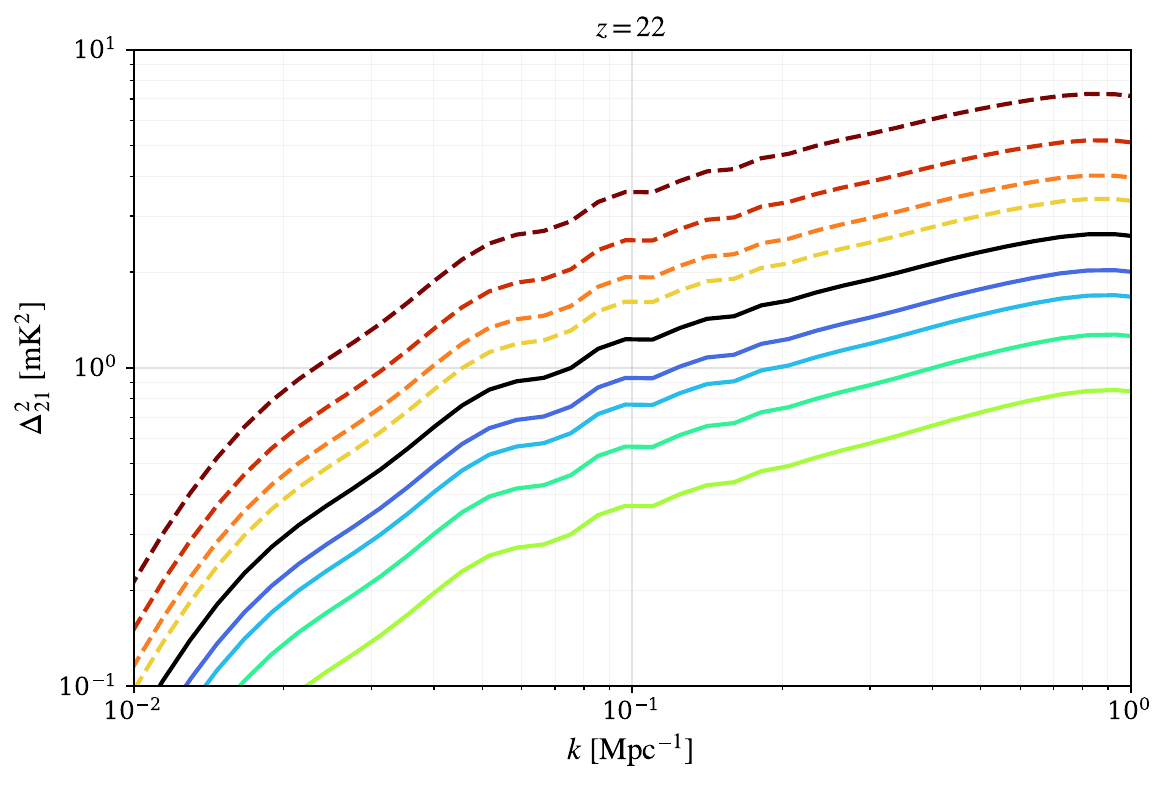}
\includegraphics[width=0.49\textwidth]{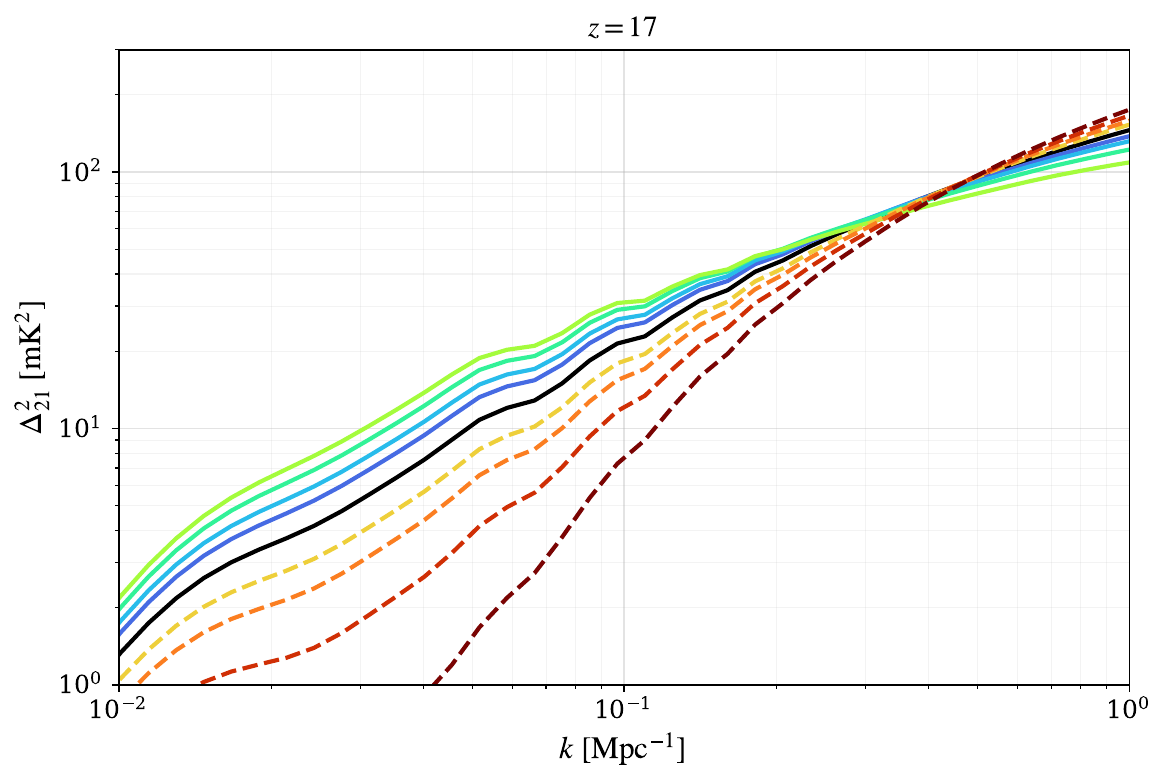}
\includegraphics[width=0.49\textwidth]{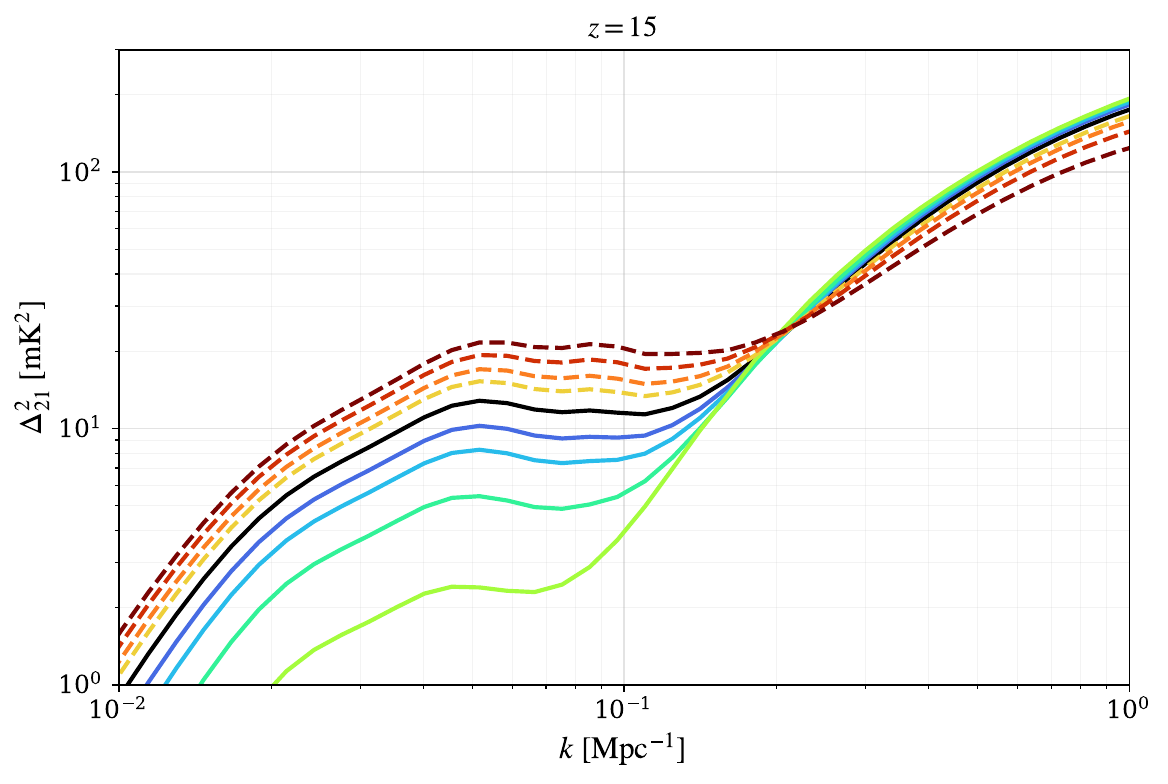}
\includegraphics[width=0.49\textwidth]{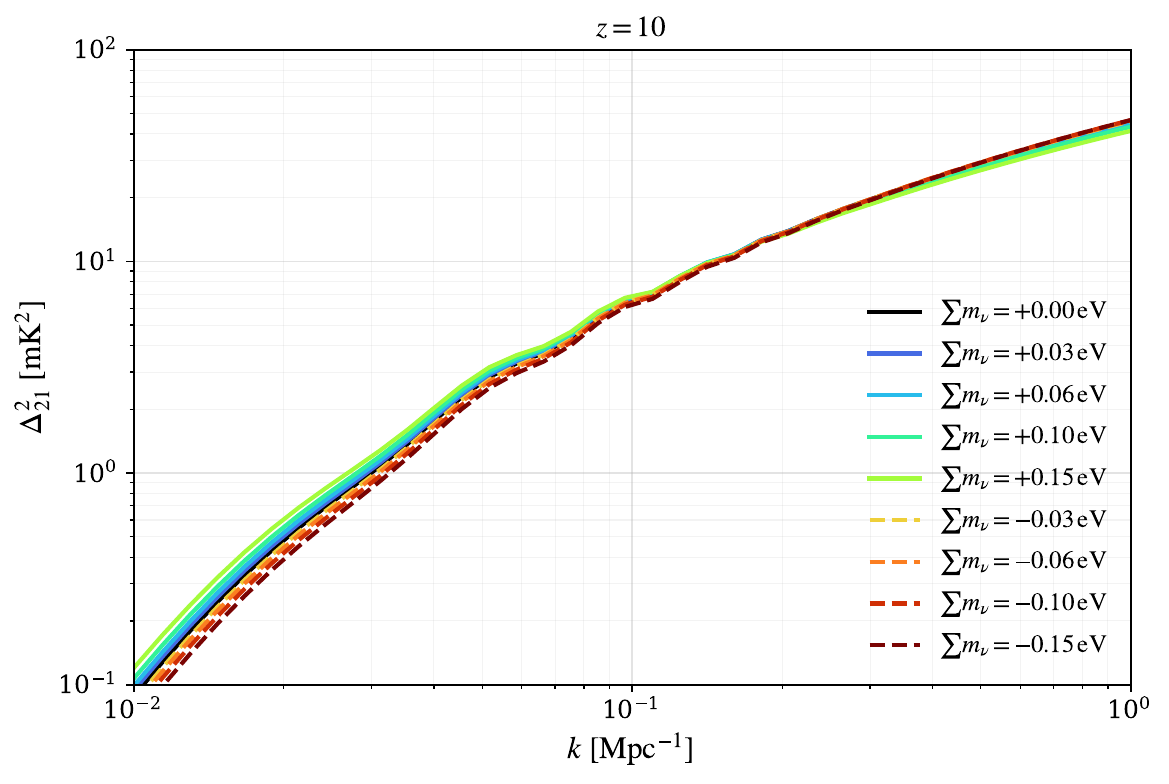}
    \caption{The 21~cm power spectra for different effective neutrino masses at four representative redshifts, $z=22, 17, 15, 10$.
Solid and dashed lines correspond to positive and negative effective neutrino masses, respectively.
}
    \label{fig: 21cm_power_spectrum}
\end{figure}

Compared with the global signal, 21~cm tomography further provides the three-dimensional spatial distribution of the 21~cm brightness temperature, 
\begin{equation}
T_{21}(\mathbf{x}, z)\propto x_{\rm HI}(\mathbf{x},z)(1+\delta_b(\mathbf{x},z))(1+z)^{1/2}\left[1-\frac{T_\gamma(z)}{T_S(\mathbf{x},z)}\right],
\label{eq: T_21}
\end{equation}
where $x_{\rm HI}(\mathbf{x},z)$ is the neutral fraction of hydrogen, $1+\delta_b(\mathbf{x},z)$ is the fractional overdensity of baryons, $T_\gamma(z)$ is the CMB temperature and $T_S(\mathbf{x},z)$ is the spin temperature.
In this work, we use the dimensionless 21~cm power spectrum as the main observable to investigate the effects of neutrino masses on the 21~cm signal.
It is defined as
\begin{equation}
\Delta_{21}^2(k,z)=\frac{k^3}{2\pi^2}P_{21}(k,z),
\end{equation}
where $P_{21}(k,z)$ is the power spectrum of $\delta T_{21}(\mathbf{x},z)\equiv T_{21}(\mathbf{x},z)-\bar{T}_{21}(z)$.
In Fig.~\ref{fig: 21cm_power_spectrum}, we show the 21~cm power spectra for different effective neutrino masses at four representative redshifts, with all other parameters fixed.

At high redshifts, as illustrated by the upper-left panel, the 21~cm power spectrum is enhanced for negative neutrino masses.
This can be attributed to the enhanced structure formation, which leads to stronger baryon density fluctuations $\delta_b$ in Eq.~\eqref{eq: T_21}.
As the redshift decreases to $z=17$, close to the trough of the global 21 cm signal in Fig.~\ref{fig: 21cm_global_signal}, the power spectrum for negative effective neutrino masses becomes lower than that for positive masses at $k\lesssim 0.1~\mathrm{Mpc}^{-1}$. 
This trend is reversed at $z=15$, when X-ray heating becomes increasingly important relative to WF coupling. 
In this regime, negative masses yield a higher power spectrum at $k\lesssim 0.1~\mathrm{Mpc}^{-1}$.
As the Universe approaches the reionization era, the differences among neutrino masses are much less apparent, as shown in the lower-right panel.
These results demonstrate that negative effective neutrino masses leave distinct redshift- and scale-dependent imprints on the 21~cm power spectrum.
We therefore use the 21~cm power spectrum as an observable to investigate the potential of future 21~cm observations in testing the negative effective neutrino-mass cosmologies favored by CMB+BAO data, with their parameter values listed in Tab.~\ref{tab: PLE_OLE_parameters}.

\subsection{21 cm Power Spectrum and Fisher Matrix}
To forecast the constraining power of future 21 cm observations, we adopt the Fisher matrix formalism based on the reduced 21 cm power spectrum $\Delta_{21}^2(k, z)$:
\begin{equation}
F_{\alpha\beta}\equiv\sum_{k,z}\frac{1}{{\rm Var}\left[\Delta_{21}^2(k, z)\right]}\frac{\partial\Delta_{21}^2(k, z)}{\partial \lambda_{\alpha}}\frac{\partial\Delta_{21}^2(k, z)}{\partial \lambda_{\beta}},
\label{eq: Fisher}
\end{equation}
where ${\rm Var}\left[\Delta_{21}^2(k, z)\right]$  is the expected uncertainty on the $\Delta_{21}^2(k, z)$ expected from a given experiment.
The parameter vector $\boldsymbol{\lambda}$ consists of five standard cosmological parameters,
\[
\left\{\omega_b, \omega_{\rm cdm}, h,n_s,\ln\left(10^{10}A_s\right)\right\},
\]
the neutrino mass parameter $\sum m_\nu$, and the astrophysical parameter vector 
\[\boldsymbol{\phi}\equiv\{f_\star, M_{\rm turn}, \zeta, L_X, \alpha_X\}\] considered in this work.
The fiducial cosmological parameter sets corresponding to the two extrapolation schemes and the reference cosmology are listed in Tab.~\ref{tab: PLE_OLE_parameters}.
We adopt a Pop II-only model for the astrophysical sources and use the same fiducial astrophysical parameters for all cosmologies considered in this work.
In the publicly available python code Zeus21~\cite{Munoz:2023kkg}, the reionization history and the corresponding optical depth are determined by the astrophysical parameters.
Therefore, we do not treat $\tau$ as an independent parameter in the Fisher matrix analysis.
We first consider a baseline scenario, in which the same fiducial astrophysical parameters are adopted for all four cosmologies.
The corresponding fiducial values and parameter definitions are listed in the second and last columns, respectively, of Tab.~\ref{tab: astro_fiducial}.
In this scenario, Zeus21 predicts an identical optical depth of approximately $\tau\simeq 0.064$ for all four cosmologies.
This value differs from the corresponding CMB+BAO-preferred optical depths listed in the last column of Tab.~\ref{tab: PLE_OLE_parameters}.
To match the optical depth predicted by Zeus21 to the corresponding CMB+BAO-preferred value for each cosmology, 
we introduce  another scenario referred to as the $\tau$-matched scenario.
In this scenario, we adjust $\zeta$ for each cosmology and keep the other astrophysical parameters unchanged.
The values of $\zeta$ are listed in the third to sixth columns of Tab.~\ref{tab: astro_fiducial}.

\begin{table*}[!htbp]
\centering
\renewcommand{\arraystretch}{1.3}
\setlength{\tabcolsep}{4pt}
\normalsize

\begin{tabular}{|c|c|c|c|c|c|p{3.2cm}|}
\hline
Parameter & Baseline & \multicolumn{4}{c|}{$\tau$-matched} & Description \\
\cline{3-6}
 & & PLE$_4$ & PLE$_9$ & OLE & Reference & \\
\hline

$f_\star$
& $0.15$ & $0.15$ & $0.15$ & $0.15$ & $0.15$
& Star formation efficiency. \\
\hline

$M_{\rm turn}\,[M_\odot]$
& $10^8$ & $10^8$ & $10^8$ & $10^8$ & $10^8$
& Turnover halo mass. \\
\hline

$\zeta$
& $75$ & $59.493$ & $59.203$ & $44.293$ & $80.354$
& Effective ionizing efficiency. \\
\hline

\shortstack{$L_X$ \\ $[{\rm erg}~{\rm s}^{-1}~M_\odot^{-1}~{\rm yr}]$}
& $3\times10^{40}$ & $3\times10^{40}$ & $3\times10^{40}$ & $3\times10^{40}$ & $3\times10^{40}$
& X-ray luminosity per unit star formation rate. \\
\hline

$\alpha_X$
& $1$ & $1$ & $1$ & $1$ & $1$
& X-ray spectral index. \\
\hline

\end{tabular}

\caption{
Definitions and fiducial values of the astrophysical parameters adopted in the Fisher matrix analysis for the baseline and $\tau$-matched scenarios.
}

\label{tab: astro_fiducial}
\end{table*}

For consistency, we use the modified version of CLASS, introduced in Sec.~\ref{Consistency of the Two Extrapolation Schemes}, for both PLE and OLE in the following 21 cm analysis. 
With these fiducial parameters, we first calculate the linear CDM~+~baryon matter power spectrum, $P_{cb}^{\rm lin}(k, z)$, and the corresponding density variance, $\sigma_{cb}(R,z)$ for each cosmology.
These quantities are then passed to Zeus21~\cite{Munoz:2023kkg}  with which the halo mass function (HMF) is calculated using the Sheth--Tormen model.
Subsequently, the 21~cm power spectrum is calculated using the resulting HMF and the phenomenological model implemented in Zeus21.
The derivatives of the 21 cm power spectrum with respect to the model parameters are then evaluated numerically using central differences around their fiducial values, while keeping all other parameters fixed.

Since HERA's compact and highly redundant array is specifically designed for 21~cm power spectrum measurements, we adopt a HERA-like configuration as a benchmark for the sensitivity of future observations.
The resulting sensitivity forecast also provides a useful reference for forecasts for other 21 cm power-spectrum experiments.
We adopt the ''optimistic'' foreground model implemented in 21cmSense~\cite{Pober:2013, Pober:2014,Murray:2024}, a total observing time of 5400 hours, and a frequency bandwidth of $8~{\rm MHz}$ per redshift bin. 
The central frequency of each band is converted into the corresponding observational redshift, which determines the redshift bins in the summation in Eq.~\ref{eq: Fisher}.
For each redshift bin, the available $k$ bins are automatically determined by 21cmSense according to the adopted observational configuration.
With these settings, we use the publicly available code 21cmSense to estimate the expected uncertainty in the 21 cm power spectrum, ${\rm Var}\left[\Delta_{21}^2(k,z)\right]$.
Finally, we calculate the Fisher matrix Eq.~\eqref{eq: Fisher} over the wavenumber window $0.1\leq k\leq1{\rm Mpc}^{-1}$ and redshift range $5\leq z\leq25$.
The forecast posterior is approximated by a Gaussian distribution centered on the fiducial parameter values $\boldsymbol{\lambda}_{\rm fid}$ listed in Tab.~\ref{tab: PLE_OLE_parameters} and Tab.~\ref{tab: astro_fiducial}, with the covariance matrix given by the inverse Fisher matrix,
\begin{equation}
P(\boldsymbol{\lambda})=\frac{\sqrt{\det F}}{(2\pi)^{N/2}}\exp\left[-\frac12(\lambda_\alpha-\lambda_{\alpha,\rm fid})F_{\alpha\beta}(\lambda_\beta-\lambda_{\beta,\rm fid})\right],
\label{eq: posterior}
\end{equation}
where $N$ is the number of parameters.
Monte Carlo (MC) samples are then generated from this multivariate Gaussian distribution and  GetDist~\cite{lewis2019getdistpythonpackageanalysing} is used to obtain the marginalized parameter constraints.

\subsection{Forecast Constraints on Neutrino Mass}
\label{Forecast Constraints on Neutrino Mass}
\begin{figure}[htbp]
\centering
\includegraphics[width=0.45\textwidth]{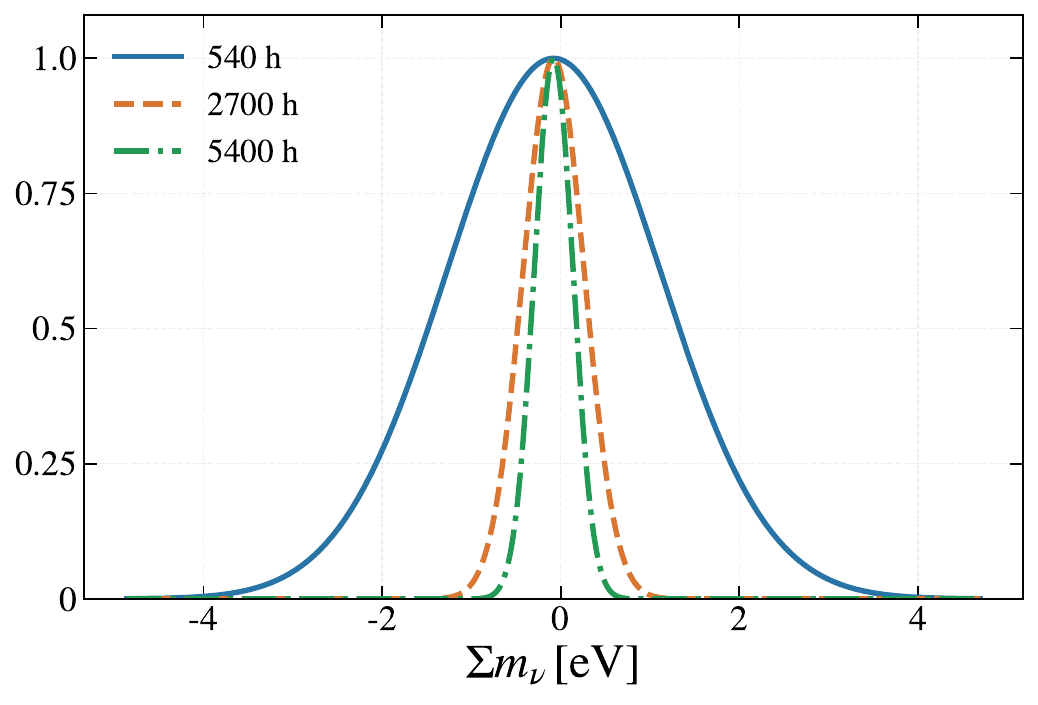}
\includegraphics[width=0.45\textwidth]{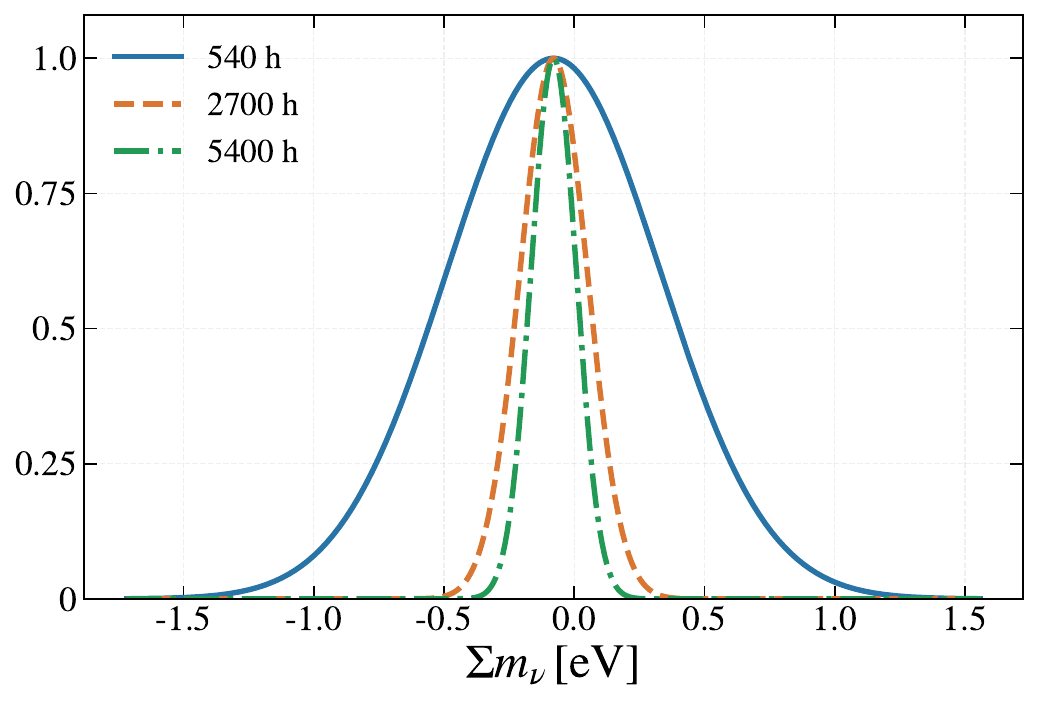}
    \caption{Marginalized 1D distributions of $\sum m_\nu$ obtained from the Fisher matrix forecasts for the OLE cosmology.
The blue solid, orange dashed, and green dash-dotted lines correspond to observing times of 540, 2700, and 5400 hours, respectively. 
The left and right panels correspond to the baseline and $\tau$-matched scenarios, respectively. All distributions are normalized to unit peak height.
}
    \label{fig: Time}
\end{figure}

Before presenting the full Fisher matrix forecast results, we examine how the observing time affects the neutrino mass constraints.
In Fig.~\ref{fig: Time}, we compare the marginalized 1D distributions of $\sum m_\nu$ for observing times of 540, 2700, and 5400 hours, using OLE as a representative cosmology.
The left and right panels correspond to the baseline and $\tau$-matched scenarios, respectively.
As the observing time increases, the uncertainties become tighter in both scenarios.
However, for observing times of 540 and 2700 hours, corresponding to approximately 0.5 and 2.5 years of observations, respectively, the marginalized uncertainties remain relatively large, indicating limited sensitivity to neutrino mass.
We therefore adopt 5400 hours, corresponding to 5 years of observations, for the following analyses.

\begin{figure}[htbp]
\centering
\includegraphics[width=0.49\textwidth]{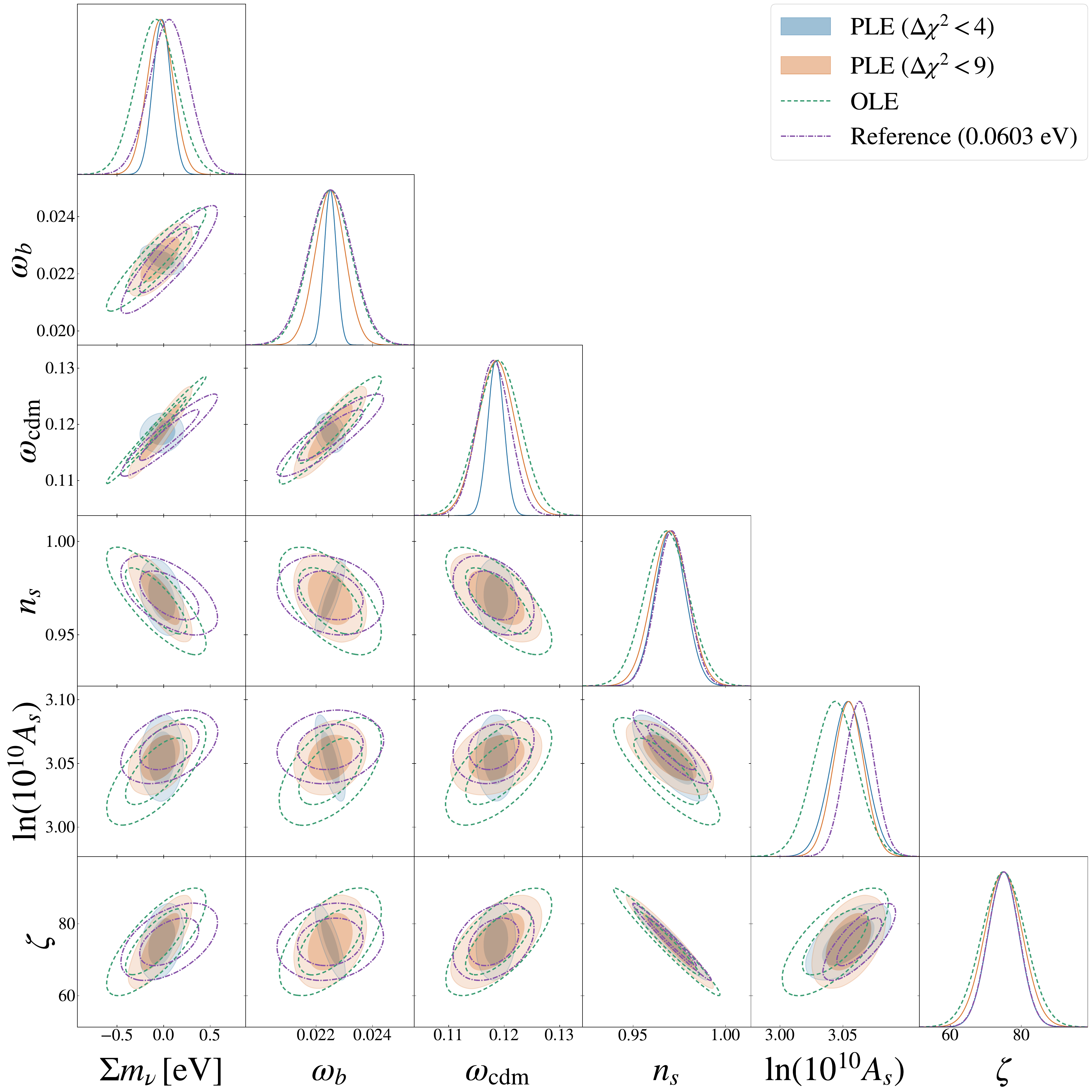}
\includegraphics[width=0.49\textwidth]{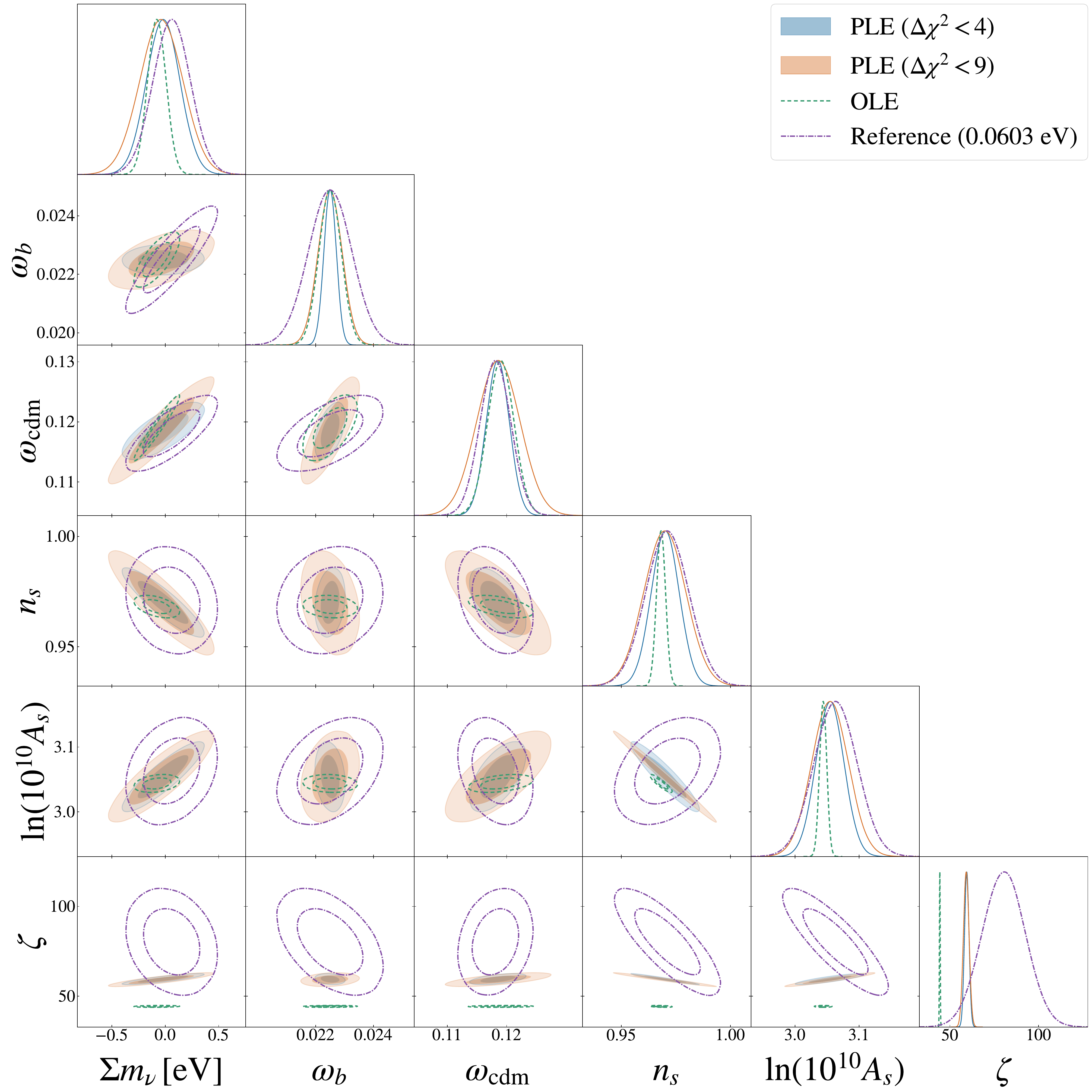}
    \caption{
The fisher matrix forecast constraints on the cosmological parameters and the effective ionizing efficiency $\zeta$ for an HERA-like 21 cm experiment. 
The left and right panels show the baseline and $\tau$-matched scenarios defined in Tab.~\ref{tab: astro_fiducial}, respectively.
The blue, orange, green, and purple distributions correspond to the PLE $(\Delta\chi^2<4)$, PLE $(\Delta\chi^2<9)$, OLE, and reference cosmologies, respectively.
The dark- and light-shaded regions represent the $1\sigma$ and $2\sigma$ confidence regions for the two PLE cases, while the dashed and dash-dotted contours represent the corresponding confidence regions for OLE and the reference cosmology, respectively.
The Fisher forecasts are based on 21~cm power spectrum over the redshift range $5\leq z\leq25$.
}
    \label{fig: optimistic_results}
\end{figure}

The full Fisher matrix forecast results are shown in Fig.~\ref{fig: optimistic_results}.
The left and right panels correspond to the baseline and $\tau$-matched scenarios, defined in Tab.~\ref{tab: astro_fiducial} respectively.
The curves in the top panel of each column show the 1D marginalized distributions, while the shaded areas show the 2D correlations between the corresponding parameters.
It is shown that the 21 cm-only forecasts yield relatively weak marginalized constraints on cosmological parameters compared with those reported by Planck 2018~\cite{Planck2018results}.

The 1D marginalized distributions of $\sum m_\nu$ show substantial overlap among the four cosmologies in both scenarios.
Moreover, the marginalized $1\sigma$ uncertainties are of order $0.1~{\rm eV}$ or larger for all four cosmologies. 
These uncertainties are relatively large compared with the differences between their fiducial neutrino mass values.
To further quantify this result, we define a separation $Z$ as
\begin{equation}
Z_{ij}=\frac{\left|(\sum m_\nu)_i-(\sum m_\nu)_j\right|}{\sigma_i(\sum m_\nu)},
\label{eq: Z_ij}
\end{equation}
where $\sigma_i(\sum m_\nu)$ is the marginalized $1\sigma$ uncertainty obtained from the Fisher matrix with cosmology $i\in\{{\rm PLE}(\Delta\chi^2<4), {\rm PLE}(\Delta\chi^2<9), {\rm OLE}, {\rm reference}$\} as the fiducial model.
Assuming that cosmology $i$ describes our universe, $Z_{ij}$ measures how much the neutrino mass of cosmology $j$ deviates from that of cosmology $i$ in units of $\sigma_i(\sum m_\nu)$.
A larger $Z_{ij}$ indicates a stronger ability of 21 cm observations to distinguish the neutrino mass $\left(\sum m_\nu\right)_j$ in cosmology $i$.
Therefore, $Z_{ij}$ is not necessarily equal to $Z_{ji}$.
\begin{table*}[!htbp]
\centering
\renewcommand{\arraystretch}{1.3}
\setlength{\tabcolsep}{6pt}
\normalsize
\begin{tabular}{|c|c|c|c|c||c|c|c|c|}
\hline
& \multicolumn{4}{c||}{Baseline} &
\multicolumn{4}{c|}{$\tau$-matched} \\
\hline
  & PLE$_4$ & PLE$_9$ & OLE & Reference &
PLE$_4$ & PLE$_9$ & OLE & Reference \\
\hline
PLE$_4$ & 0 & 0.1793 & 0.6220 & 0.8331 &
0 & 0.1096 & 0.3799 & 0.5089 \\
\hline
PLE$_9$ & 0.1228 & 0 & 0.3032 & 0.6934 &
0.0850 & 0 & 0.2098 & 0.4799 \\
\hline
OLE & 0.2737 & 0.1947 & 0 & 0.6402 &
0.6837 & 0.4865 & 0 & 1.5994 \\
\hline
Reference & 0.3782 & 0.4596 & 0.6605 & 0 &
0.4573 & 0.5557 & 0.7987 & 0 \\
\hline
\end{tabular}
\caption{
Separation $Z_{ij}$ between the neutrino mass values of different cosmologies in the baseline (left) and $\tau$-matched (right) scenarios.
The subscripts 4 and 9 in PLE$_4$ and PLE$_9$ denote $\Delta\chi^2<4$ and $\Delta\chi^2<9$, respectively.
The rows correspond to the assumed true cosmologies $i$, while the columns correspond to the tested cosmologies $j$. 
The separations are calculated using 21~cm power spectrum over the redshift range $5\leq z\leq25$.
}
\label{tab: Z_matrix}
\end{table*}
The resulting $Z_{ij}$ values are listed in Tab.~\ref{tab: Z_matrix}.
Most of the values are smaller than 1, quantitatively confirming the substantial overlap of the 1D marginalized neutrino mass distributions shown in Fig.~\ref{fig: optimistic_results}.
In particular, the two PLE cases are difficult to distinguish, with $Z_{ij}$ below 0.2 in both scenarios.
The separations between PLE and OLE also remain below 1, indicating that the differences in the $\sum m_\nu$ predicted by the two extrapolation schemes cannot be distinguished by 21 cm probes alone.

An exception occurs in the $\tau$-matched scenario, where $Z_{\rm OLE,Reference}$ increases from 0.6402 in the baseline scenario to 1.5994.
This value corresponds to the separation between the green OLE curve in the right panel of Fig.~\ref{fig: optimistic_results} and the reference neutrino mass value of $0.0603~{\rm eV}$.
The enhancement is associated with the reduction in the fiducial effective ionizing efficiency $\zeta$ for OLE, from 75 in the baseline scenario to 44.293 in the $\tau$-matched scenario.
\begin{figure}[htbp]
\centering
\includegraphics[width=0.9\textwidth]{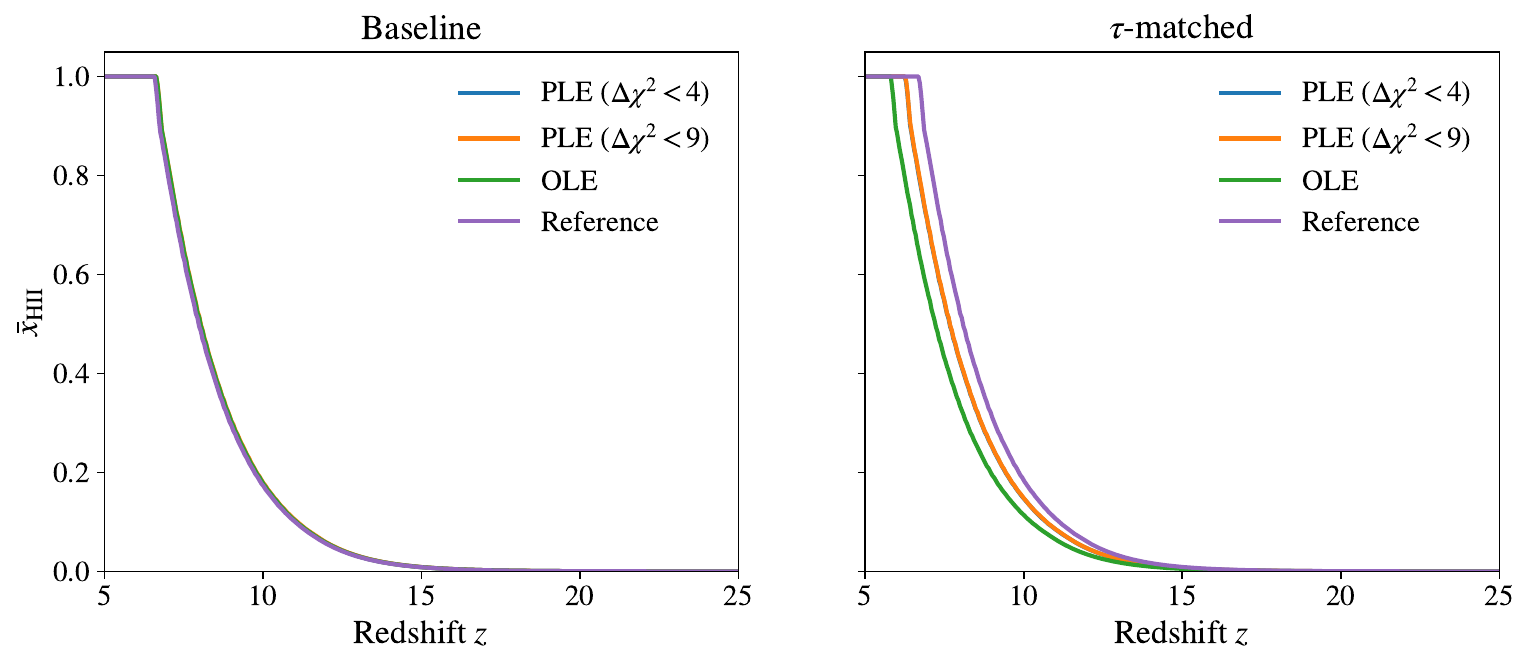}
    \caption{Evolution of the mean ionized hydrogen fraction $\bar{x}_{\rm HII}$ as a function of redshift for the four cosmologies.
The left and right panels correspond to the baseline and $\tau$-matched scenarios, respectively.
The shifts in the reionization histories between the two scenarios are mainly driven by changes in the fiducial ionizing efficiency $\zeta$.
The four reionization histories nearly overlap in the left panel, while only the two PLE cases nearly overlap in the right panel.
}
    \label{fig: reionization_history}
\end{figure}
A smaller $\zeta$ in the OLE case leads to a delayed reionization, as shown in Fig.~\ref{fig: reionization_history}.
The delayed reionization leaves a higher abundance of neutral hydrogen and thus enhances the 21 cm signal over the most sensitive redshift range. 
The marginalized constraint on $\zeta$ consequently becomes tighter in the $\tau$-matched scenario, as shown in the bottom-right subplot of Fig.~\ref{fig: optimistic_results}.
Since $\zeta$ is correlated with $\sum m_\nu$, as shown in the bottom-left subplot of the left panel of Fig.~\ref{fig: optimistic_results}, the tighter constraint on $\zeta$ helps break their degeneracy, thereby tightening the marginalized constraint on $\sum m_\nu$ for OLE.
For the two PLE cases, however, the marginalized uncertainties on $\sum m_\nu$ become larger in the $\tau$-matched scenario, despite the improved constraints on $\zeta$.

Overall, all the separation values $Z_{ij}$ remain below 2 in both scenarios.
This suggests that, after marginalizing over the other parameters, 21 cm probes alone have limited ability to distinguish the negative effective neutrino-mass values favored by CMB+BAO from the reference positive value.

\begin{figure}[htbp]
\centering
\includegraphics[width=0.49\textwidth]{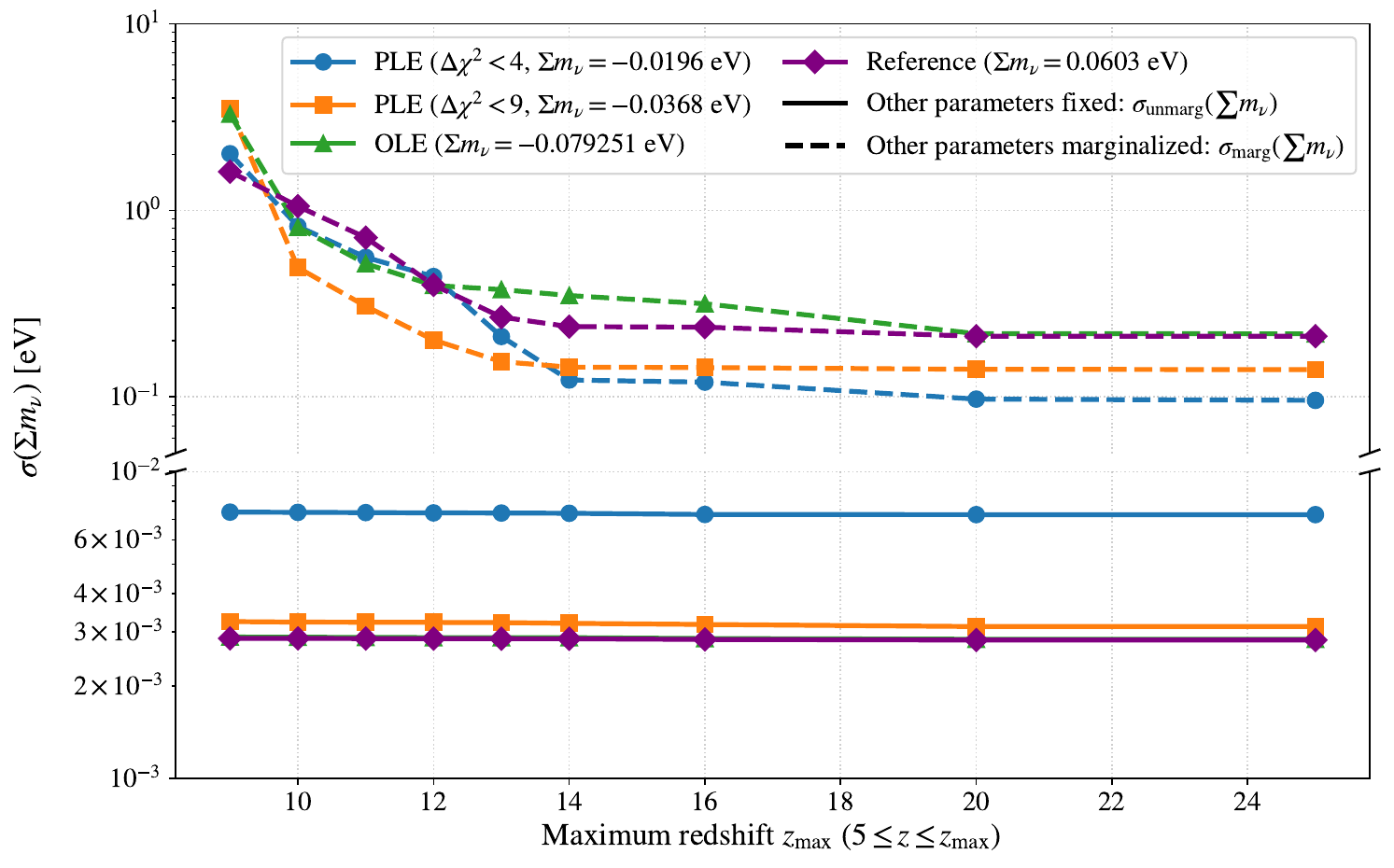}
\includegraphics[width=0.49\textwidth]{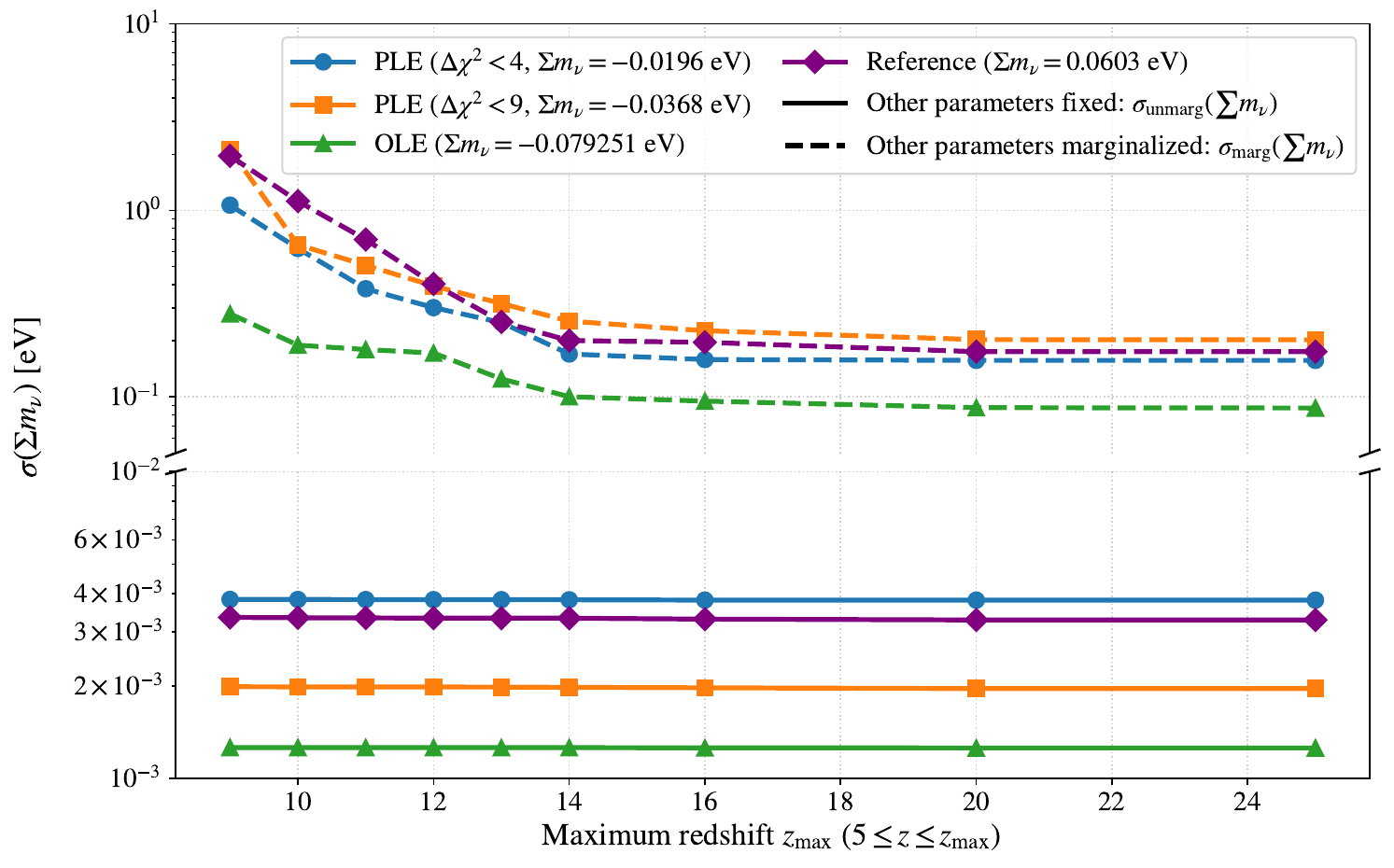}
    \caption{Marginalized and fixed-parameter uncertainties on $\sum m_\nu$, as defined in Eqs.~\eqref{eq: marg_constraint} and \eqref{eq: fixed_constraint}, as functions of the maximum redshift $z_{\max}$ for the four fiducial cosmologies.
The left and right panels correspond to the baseline and $\tau$-matched scenarios, respectively.
The blue circles, orange squares, green triangles, and purple diamonds represent the PLE ($\Delta\chi^2<4$), PLE ($\Delta\chi^2<9$), OLE, and reference cosmologies, respectively. 
The dashed lines show the marginalized uncertainties, while the solid lines show the fixed-parameter uncertainties, with all other parameters fixed at their fiducial values listed in Tab.~\ref{tab: PLE_OLE_parameters} and Tab.~\ref{tab: astro_fiducial}.
Note that the vertical axes are broken to display the marginalized and fixed-parameter uncertainties on different scales.
}
    \label{fig: degeneracy_fixormarg_z}
\end{figure}

This limited separation arises from parameter degeneracies.
To quantify the impact of these degeneracies on 21~cm constraints,  we compare the marginalized and fixed-parameter uncertainties on neutrino mass.
The marginalized uncertainty on a parameter $\lambda_\alpha$ is defined as
\begin{equation}
\sigma_{{\rm marg}}(\lambda_\alpha)\equiv\sqrt{(F^{-1})_{\alpha\alpha}},
\label{eq: marg_constraint}
\end{equation}
where $(F^{-1})_{\alpha\alpha}$ is the diagonal element of the inverse Fisher matrix, with $F$ defined in Eq.~\eqref{eq: Fisher}.
The fixed-parameter uncertainty is obtained by fixing all other parameters at their fiducial values and is given by
\begin{equation}
\sigma_{{\rm unmarg}}(\lambda_\alpha)\equiv\frac{1}{\sqrt{F_{\alpha\alpha}}}.
\label{eq: fixed_constraint}
\end{equation}
Parameter degeneracy occurs when changes in different parameters produce similar effects on the 21~cm power spectrum. 
When other parameters are allowed to vary, their effects can partially compensate for changes in neutrino mass, leading to a larger marginalized uncertainty.
To further examine how the redshift coverage of 21 cm observations affects parameter degeneracies, we introduce the maximum redshift $z_{\max}$.
The redshift range included in the summation of the Fisher matrix defined in Eq.~\eqref{eq: Fisher} is then given by $5\leq z\leq z_{\max}$.
The marginalized and fixed-parameter uncertainties as functions of $z_{\max}$ are shown in Fig.~\ref{fig: degeneracy_fixormarg_z} as dashed and solid lines, respectively.
The left panel corresponds to the baseline scenario, while the right panel shows the $\tau$-matched scenario.
Since reionization is nearly complete at $z\lesssim7$, the 21 cm signal becomes negligible in this redshift range. We therefore restrict the displayed results to $z_{\max}>9$.

In both scenarios, the fixed-parameter uncertainties on $\sum m_\nu$ are of order $10^{-3}~{\rm eV}$.
In contrast, the marginalized uncertainties are much larger, reaching $0.1$–$1~{\rm eV}$ in both scenarios.
This large difference indicates that parameter degeneracies significantly weaken the constraining power of 21 cm probes on neutrino mass.
Furthermore, despite the reduction in the amount of observational data as $z_{\max}$ decreases, the fixed-parameter uncertainties vary by only a few percent over the displayed redshift range.
This suggests that the additional high-redshift data contribute little to the neutrino mass sensitivity when the other parameters are fixed.
However, the marginalized uncertainties decrease significantly as $z_{\max}$ increases.
This means that including high-redshift 21 cm power spectrum data can help break parameter degeneracies and improve the neutrino mass constraints,  although this improvement gradually saturates.
\begin{figure}[htbp]
\centering
\includegraphics[width=0.95\textwidth]{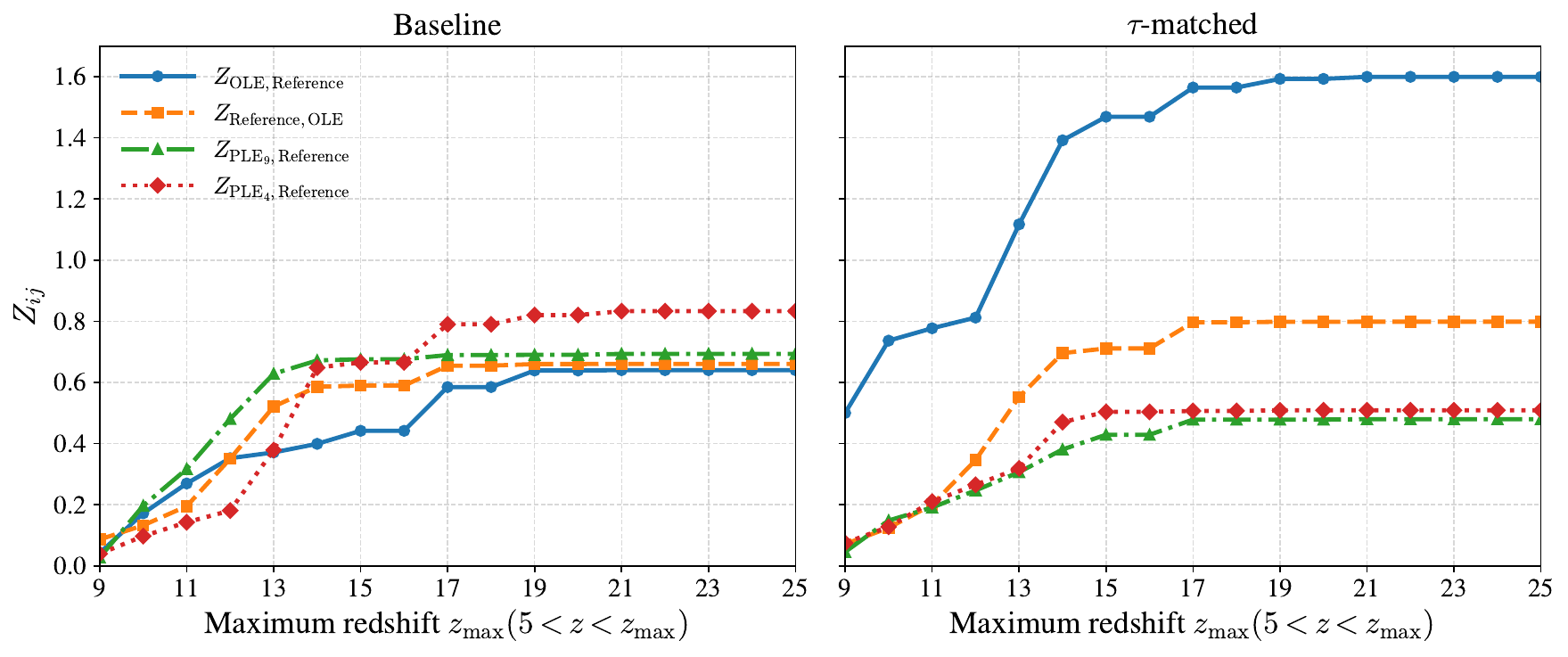}
    \caption{Separations $Z_{ij}$ as a function of $z_{\rm max}$ for selected pairs of cosmologies in the baseline (left) and $\tau$-matched (right) scenarios.
The points at $z_{\rm max}=25$ correspond to the separation values reported in Tab.~\ref{tab: Z_matrix}.
}
    \label{fig: Z_zmax}
\end{figure}
To further investigate the impact of including higher-redshift data, we plot selected separations $Z_{ij}$ as functions of $z_{\max}$ in Fig.~\ref{fig: Z_zmax}.
All the separations increase as $z_{\max}$ increases, indicating an improved ability to alleviate parameter degeneracies and thus distinguish different cosmologies.
However, this improvement gradually saturates at $z_{\max}\sim14$--18.

Finally, we highlight an interesting phenomenon.
In the left panel  of  Fig.~\ref{fig: degeneracy_fixormarg_z}, the solid lines show that $\sigma_{\rm unmarg}(\sum m_\nu)$ is significantly larger for PLE ($\Delta\chi^2<4$) than for the other three cases, which yield nearly identical values.
Since all other parameters are fixed in calculating $\sigma_{\rm unmarg}$, this difference is not caused by parameter degeneracies.
Moreover, the four cosmologies have nearly identical reionization histories, as shown in Fig.~\ref{fig: reionization_history}.
Therefore, the difference should mainly arise from the sensitivity of the 21~cm power spectrum to neutrino mass.
In particular, the fiducial neutrino mass determines the redshift $z_{\rm nr}$, at which neutrinos become nonrelativistic.
For three degenerate neutrino species, this redshift can be estimated as
\begin{equation}
1+z_{\rm nr}\simeq 630\left(\frac{|\sum m_\nu|}{1{\rm eV}}\right),
\end{equation}
where the absolute value is consistent with the phenomenological effective neutrino model defined in Eq.~\eqref{eq: effective neutrino}.
This gives $z_{\rm nr}\simeq 11$, $22$, $49$, and $37$ for PLE ($\Delta\chi^2<4$), PLE ($\Delta\chi^2<9$), OLE, and the reference cosmology, respectively.
Therefore, neutrinos in PLE ($\Delta\chi^2<4$) remain relativistic until a much lower redshift than in the other three cases.
The effects of neutrino mass on the structure formation and 21 cm signal thus emerge later and are weaker.
Consequently, this case yields a larger $\sigma_{\rm unmarg}(\sum m_\nu)$.


\subsection{Impact of External Cosmological Constraints}
\label{Impact of External Cosmological Constraints}
As discussed above, 21 cm observations alone have limited ability to distinguish among the neutrino mass cosmologies considered here, largely due to parameter degeneracies.
Other independent cosmological probes, such as CMB and BAO, can provide complementary constraints on these parameters, helping to break the degeneracies.
Therefore, to investigate the potential impact of such external information, we modify the 21 cm Fisher matrix in Eq.~\eqref{eq: Fisher} by adding the contribution from the Gaussian priors, such that
\begin{equation}
F_{\alpha\beta}^{\rm total}\equiv F_{\alpha\beta}+F_{\alpha\beta}^{\rm prior},
\label{eq: Fisher_prior}
\end{equation}
where the prior Fisher matrix is given by
\begin{equation}
F_{\alpha\beta}^{\rm prior}\equiv
\left\{
\begin{array}{ll}
\dfrac{1}{\sigma_{\alpha,\rm prior}^{2}}
& \quad\quad\quad \text{when } \alpha=\beta
\text{ and }
\lambda_\alpha\in S,\\
0
& \quad\quad\quad \text{otherwise}.
\end{array}
\right.
\end{equation}
Here, Gaussian priors are imposed on $S\equiv\left\{\omega_b, \omega_{\rm cdm}, \ln(10^{10}A_s), n_s\right\}$.
For simplicity, we neglect correlations among these parameters and treat their priors as independent.
Given the Hubble tension, we do not impose an external prior on $h$.
We first adopt the marginalized $1\sigma$ uncertainties from Planck 2018 data combined with BAO measurements~\cite{Planck2018results} as the widths of the Gaussian priors, as listed in Tab.~\ref{tab:prior_widths}.
\begin{table}[htbp]
\centering
\renewcommand{\arraystretch}{1.5}
\begin{tabular}{cc}
\hline
Parameter $\lambda_\alpha$ & $\sigma_{\alpha,\rm prior}^{\rm Planck~2018~+~BAO}$ \\
\hline
$\omega_b$ & 0.00014 \\
$\omega_{\rm cdm}$ & 0.00091 \\
$\ln(10^{10}A_s)$ & 0.014 \\
$n_s$ & 0.0038 \\
\hline
\end{tabular}
\caption{Gaussian prior widths adopted in the Fisher analysis, based on the marginalized $1\sigma$ constraints from Planck~2018~+~BAO~\cite{Planck2018results}.}
\label{tab:prior_widths}
\end{table}
Following the procedure described in Sec.~\ref{Forecast Constraints on Neutrino Mass}, we calculate the separation values $Z_{ij}$ using the marginalized uncertainties obtained from $F_{\alpha\beta}^{\rm total}$. 
The results are listed in Tab.~\ref{tab: Z_matrix_reference}.
\begin{table*}[!htbp]
\centering
\renewcommand{\arraystretch}{1.3}
\setlength{\tabcolsep}{6pt}
\normalsize
\begin{tabular}{|c|c|c|c|c||c|c|c|c|}
\hline
& \multicolumn{4}{c||}{Baseline} &
\multicolumn{4}{c|}{$\tau$-matched} \\
\hline
& PLE$_4$ & PLE$_9$ & OLE & Reference &
PLE$_4$ & PLE$_9$ & OLE & Reference \\
\hline
PLE$_4$ & 0 & 0.1923 & 0.6669 & 0.8932 &
0 & 0.2332 & 0.8087 & 1.0833 \\
\hline
PLE$_9$ & 0.4039 & 0 & 0.9969 & 2.2801 &
0.2933 & 0 & 0.7239 & 1.6557 \\
\hline
OLE & 1.4879 & 1.0589 & 0 & 3.4808 &
2.0399 & 1.4517 & 0 & 4.7723 \\
\hline
Reference & 1.3342 & 1.6214 & 2.3303 & 0 &
1.3448 & 1.6343 & 2.3487 & 0 \\
\hline
\end{tabular}
\caption{
Separation $Z_{ij}$ between the neutrino mass values of different cosmologies in the baseline (left) and $\tau$-matched (right) scenarios, obtained with Gaussian priors on cosmological parameters.
The prior widths are adopted from the marginalized $1\sigma$ constraints from Planck~2018~+BAO.
The subscripts 4 and 9 in PLE$_4$ and PLE$_9$ denote $\Delta\chi^2<4$ and $\Delta\chi^2<9$, respectively.
The rows correspond to the assumed true cosmologies $i$, while the columns correspond to the tested cosmologies $j$.
The separations are calculated using 21~cm power spectrum over the redshift range $5\leq z\leq25$.
}
\label{tab: Z_matrix_reference}
\end{table*}
Compared with the 21~cm only results in Tab.~\ref{tab: Z_matrix}, all off-diagonal $Z_{ij}$ values increase when Gaussian priors in Tab.~\ref{tab:prior_widths} are imposed.
In particular, both $Z_{\rm OLE,Reference}$ and $Z_{\rm Reference,OLE}$ exceed 2 in both scenarios.
This indicates that incorporating CMB+BAO observations could enhance the ability of future 21~cm observations to distinguish the OLE and reference cosmologies.
However, most of the $Z_{ij}$ values involving the PLE cosmologies remain below 2, even with the additional external constraints.

The improvement in $Z_{ij}$ can be understood in terms of parameter degeneracies.
Changes in the four cosmological parameters can partially compensate for the effects of neutrino mass on the 21~cm power spectrum.
The Gaussian priors in Tab.~\ref{tab:prior_widths} restrict these compensating variations, and therefore narrower priors could further reduce their impact.

\begin{figure}[htbp]
\centering
\includegraphics[width=0.95\textwidth]{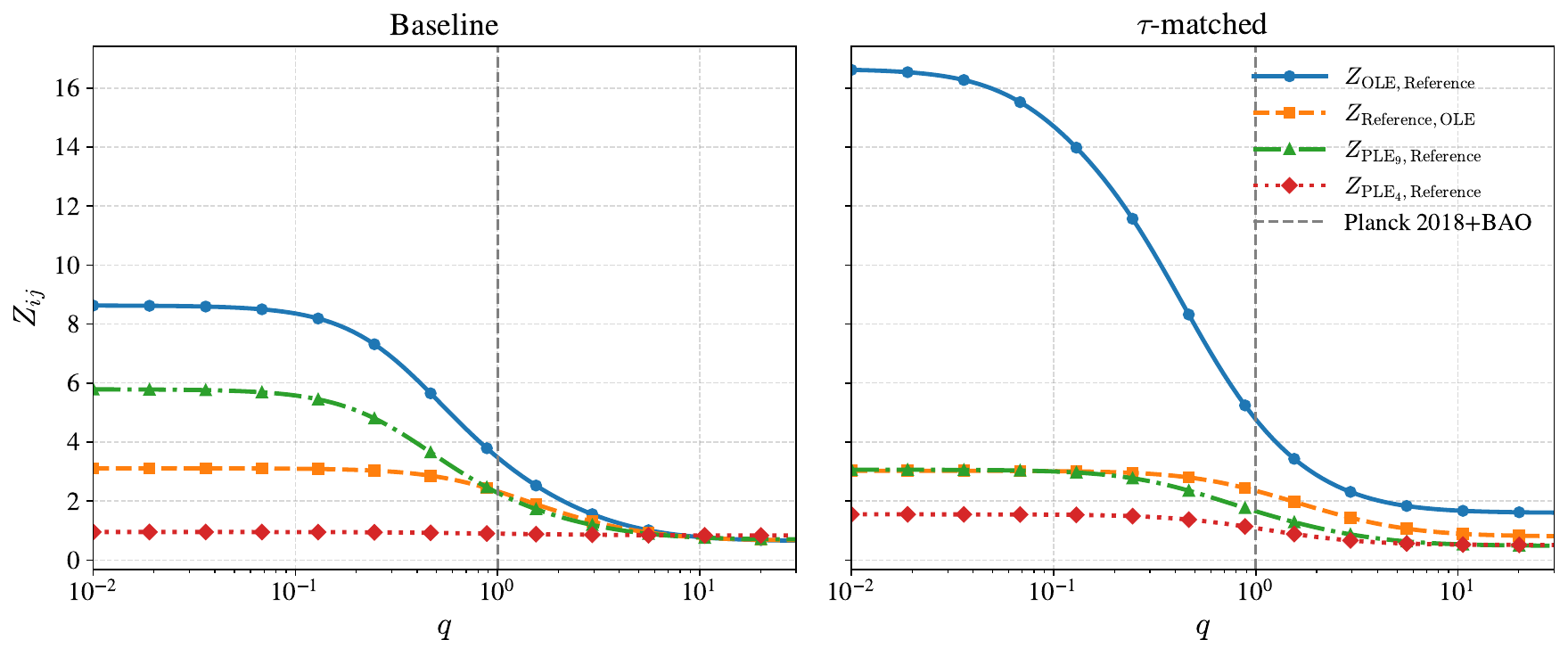}
    \caption{Separation $Z_{ij}$ as a function of the scaling factor $q$ for selected pairs of cosmologies in the baseline (left) and $\tau$-matched (right) scenarios.
The Gaussian prior widths are scaled according to $q$, as defined in Eq.~\eqref{eq: scale_q}.
The vertical dashed lines indicate $q=1$, corresponding to the Planck~2018~+~BAO priors $\sigma_{\alpha, \rm prior}^{\rm Planck 2018 + BAO}$ listed in Tab.~\ref{tab:prior_widths}.
The separations are calculated using 21~cm power spectrum over the redshift range $5\leq z\leq25$.
}
    \label{fig: Z_q}
\end{figure}

Looking ahead, future CMB and BAO observations are expected to provide increasingly stringent constraints on cosmological parameters. 
To quantify how much these improvements could increase the separation $Z_{ij}$, we introduce a scaling factor $q$ for the Gaussian prior widths,
\begin{equation}
\sigma_{\alpha,\rm prior}(q)\equiv q\sigma_{\alpha,\rm prior}^{\rm Planck 2018 + BAO}\quad\quad\quad\quad \lambda_\alpha\in S.
\label{eq: scale_q}
\end{equation}
Here, $q=1$ corresponds to the Planck 2018+BAO constraints, while $q<1$ represents tighter external priors.
The resulting $Z_{ij}$ values as functions of $q$ for selected pairs of cosmologies are shown in Fig.~\ref{fig: Z_q}.
As shown in Fig.~\ref{fig: Z_q}, all the separation values $Z_{ij}$ increase as $q$ decreases,  indicating better discrimination between different cosmologies.
This improvement arises because narrower Gaussian priors reduce the degeneracies between neutrino mass and the four cosmological parameters.
As these degeneracies become weaker, the improvement in $Z_{ij}$ becomes smaller, and eventually $Z_{ij}$ approaches a constant.

Future CMB and BAO observations~\cite{Abazajian:2019eic, Sehgal:2019ewc, SimonsObservatory:2025wwn, Euclid:2024yrr, Zhao:2024alp} are expected to provide increasingly tighter constraints on cosmological parameters~\cite{MacInnis:2023vif}.
Motivated by this, we consider a representative scenario with $q=0.4$.
The resulting separation values are listed in Tab.~\ref{tab: Z_matrix_4more_accurate}.
It is shown that $Z_{\rm OLE, Reference}$ reaches 6.1059 and 9.1418 in the baseline and $\tau$-matched scenarios, respectively,  while $Z_{\rm Reference,OLE}$ reaches 2.9181 and 2.8528.
These separation values, which are close to or exceed 3, indicate that future 21~cm observations could distinguish the OLE and reference cosmologies if the four cosmological parameters $\omega_b$, $\omega_{\rm cdm}$, $A_s$, and $n_s$ are more precisely measured by CMB and BAO.

According to the definition of $Z_{ij}$ in Eq.~\eqref{eq: Z_ij}, these large values arise from two factors. 
First, the relatively large difference in the fiducial neutrino masses between OLE and the reference cosmology leads to a larger numerator.
Second, the reduction in parameter degeneracies leads to a smaller marginalized uncertainty on $\sum m_\nu$, and thus a smaller denominator.
The role of the denominator is further illustrated by comparing $Z_{{\rm PLE}_4,{\rm Reference}}$ in the two scenarios.
For ${\rm PLE}_4$, the Gaussian priors on the four cosmological parameters have small impact on the marginalized neutrino mass constraint in the baseline scenario.
In the $\tau$-matched scenario, however, these priors more effectively reduce the parameter degeneracies, leading to tighter neutrino mass constraints.
The denominator of $Z_{{\rm PLE}_4,{\rm Reference}}$ thus remains larger in the baseline scenario, while it is reduced in the $\tau$-matched scenario, with the numerator unchanged between the two scenarios.
Consequently, imposing the Gaussian priors leads to only a slight increase in $Z_{{\rm PLE}_4,{\rm Reference}}$ from $0.8331$ to $0.9270$ in the baseline scenario, but a much larger increase from $0.5089$ to $1.4093$ in the $\tau$-matched scenario.
Further support for this interpretation is provided in Appendix~\ref{app: degeneracy}, particularly in the left panels of Fig.~\ref{fig: degeneracy_R_z_app}.

More generally, comparing Tabs.~\ref{tab: Z_matrix} and \ref{tab: Z_matrix_4more_accurate}, we find that all the separation values $Z_{ij}$ between the negative effective neutrino mass cosmologies and the reference cosmology increase when the four cosmological parameters are more precisely measured, in both scenarios.

\begin{table*}[!htbp]
\centering
\renewcommand{\arraystretch}{1.3}
\setlength{\tabcolsep}{6pt}
\normalsize

\begin{tabular}{|c|c|c|c|c||c|c|c|c|}
\hline
& \multicolumn{4}{c||}{Baseline} &
\multicolumn{4}{c|}{$\tau$-matched} \\
\hline
& PLE$_4$ & PLE$_9$ & OLE & Reference &
PLE$_4$ & PLE$_9$ & OLE & Reference \\
\hline
PLE$_4$ & 0 & 0.1996 & 0.6921 & 0.9270 &
0 & 0.3034 & 1.0521 & 1.4093 \\
\hline
PLE$_9$ & 0.7010 & 0 & 1.7302 & 3.9575 &
0.4385 & 0 & 1.0822 & 2.4754 \\
\hline
OLE & 2.6100 & 1.8574 & 0 & 6.1059 &
3.9077 & 2.7809 & 0 & 9.1418 \\
\hline
Reference & 1.6708 & 2.0304 & 2.9181 & 0 &
1.6334 & 1.9850 & 2.8528 & 0 \\
\hline
\end{tabular}

\caption{Separation $Z_{ij}$ between the neutrino mass values of different cosmologies in the baseline (left) and $\tau$-matched (right) scenarios, obtained with Gaussian priors on cosmological parameters.
The prior widths are obtained by setting $q=0.4$, where $q$ is defined in Eq.~\eqref{eq: scale_q}.
The subscripts 4 and 9 in PLE$_4$ and PLE$_9$ denote $\Delta\chi^2<4$ and $\Delta\chi^2<9$, respectively.
The rows correspond to the assumed true cosmologies $i$, while the columns correspond to the tested cosmologies $j$.
The separations are calculated using 21~cm power spectrum over the redshift range $5\leq z\leq25$.
}
\label{tab: Z_matrix_4more_accurate}
\end{table*}

However, several other separation values in Tab.~\ref{tab: Z_matrix_4more_accurate} remain below $2$.
Moreover, several separation values remain relatively small even in the limit $q\to0$, as shown in Fig.~\ref{fig: Z_q}. 
This suggests that degeneracies between neutrino mass and the remaining free parameters, including $h$ and astrophysical parameters, still play an important role in limiting the neutrino mass constraints.
More discussions on parameter degeneracies are presented in Appendix~\ref{app: degeneracy}.

\section{Conclusions}
\label{Conclusions}
In this work, we have investigated the potential of future 21~cm observations to distinguish cosmologies with negative effective neutrino masses favored by CMB~+~BAO data from a reference cosmology with $\sum m_\nu=0.06~{\rm eV}$.

We considered two phenomenological extrapolation schemes, the Profile-Likelihood Extrapolation (PLE) and the Observable-Level Extrapolation (OLE), and yielded three cosmologies with negative effective neutrino masses, as listed in Tab.~\ref{tab: PLE_OLE_parameters}.
Although their cosmological parameters differ, their predictions for CMB and BAO observables are in good agreement with each other.

We then calculated the 21~cm power spectra and used the Fisher matrix formalism to forecast the sensitivity under the baseline and $\tau$-matched scenarios.
We found that 21~cm observations alone have limited ability to distinguish these cosmologies.
This limitation is mainly due to degeneracies between the neutrino mass and other cosmological and astrophysical parameters.
Including higher redshift 21~cm observations can help reduce these parameter degeneracies and improve the marginalized neutrino mass constraints, although this improvement gradually saturates at $z_{\max}\sim14-18$.

We further found that imposing independent external Gaussian priors on four cosmological parameters, based on Planck 2018 + BAO constraints, can further alleviate these degeneracies and improve the ability of 21~cm observations to distinguish the cosmologies.

In the future, CMB experiments, including CMB-S4~\cite{Abazajian:2019eic}, SO~\cite{SimonsObservatory:2025wwn} and CMB-HD~\cite{Sehgal:2019ewc}, together with BAO surveys such as the Euclid mission~\cite{Euclid:2024yrr} and MUST~\cite{Zhao:2024alp} are expected to provide tighter constraints on cosmological parameters.
Such constraints could further alleviate parameter degeneracies in 21~cm observations, thereby improving their potential to constrain the sum of neutrino masses in cosmology.

\acknowledgments
W. Liao is supported by National Natural Science Foundation of China under the grant No. 11875130.

\appendix

\section{Effective Neutrino Free Streaming}
\label{Effective Neutrino Free Streaming}
In our study, we introduced a phenomenological description of effective negative neutrino masses in modified CLASS, as defined in Eq.~\eqref{eq: effective neutrino}.

At background level, additional ultrarelativistic component is introduced
 in our model compensates for the negative neutrino energy density in the relativistic limit.
This preserves the standard radiation density and hence the expansion history during radiation domination.
Once neutrinos become nonrelativistic, however, the neutrino effective energy density contributes to the matter density.
In this limit, the Hubble expansion rate is 
\begin{equation}
H(a) = H_0\sqrt{\Omega_m a^{-3} + \Omega_\Lambda},
\label{eq: Hubble}
\end{equation}
where $\Omega_i\equiv\omega_i/h^2\equiv\rho_i/\rho_{{\rm crit},0}$ and $\rho_{{\rm crit},0}\equiv3H_0^2/8\pi G$.
The matter density is given by $\omega_m\equiv\omega_c+\omega_b+\omega_{\nu, \rm eff}$, where $\omega_{\nu,\rm eff}\simeq\frac{\sum m_\nu}{93.14\rm eV}$ is the effective neutrino density, which becomes negative for $\sum m_\nu<0$.
At fixed $\Omega_m$, $h$, and $\omega_b$, an increase in $\sum m_\nu$ must be compensated by a decrease in $\omega_{\rm cdm}$ to preserve the total matter density and thus the expansion rate.
This leads to an approximate degeneracy between the two parameters at the background level.

At perturbation level, the energy-momentum conservation equations for neutrinos in the conformal Newtonian gauge are given by 
\begin{equation}
\dot{\delta_\nu} = -(1+w_\nu)(\theta_\nu-3\dot\phi)-3\frac{\dot a}{a}\left(\frac{\delta P_\nu}{\delta\rho_\nu}-w_\nu\right)\delta_\nu,
\label{eq: conti}
\end{equation}
\begin{equation}
\dot\theta_\nu = -\frac{\dot a}{a}(1-3w_\nu)\theta_\nu - \frac{\dot{w}_\nu}{1+w_\nu}\theta_\nu+\frac{\delta P_\nu/\delta\rho_\nu}{1+w_\nu}k^2\delta_\nu - k^2\sigma_\nu+k^2\psi.
\label{eq: eular}
\end{equation}
Here $\delta_\nu\equiv\delta\rho_\nu/\bar{\rho}_\nu$ is the neutrino density contrast, $w_\nu\equiv\bar P_\nu/\bar\rho_\nu$ is the equation of state , $\theta_\nu$ is the velocity divergence, and $\sigma_\nu$ is the anisotropic stress. 
$\phi$ and $\psi$ are the scalar metric perturbations, $k$ is the comoving wavenumber, and overdots denote derivatives with respect to conformal time.
The perturbed energy density, pressure, energy flux, and shear stress in $k$-space can be expressed in terms of the multipoles of the neutrino distribution function,
\begin{equation}
\delta\rho_\nu = 4\pi a^{-4}\int y^2dy\epsilon_{\nu\rm, eff}f_{\nu0}(y)\Psi_0
\label{eq: perturbation_1}
\end{equation}
\begin{equation}
\delta P_\nu = \frac{4\pi}{3} a^{-4}\int y^2dy\frac{y^2}{\epsilon_{\nu\rm, eff}}f_{\nu0}(y)\Psi_0
\label{eq: perturbation_2}
\end{equation}
\begin{equation}
\left(\bar{\rho}_\nu+\bar{P}_\nu\right)\theta_\nu = 4\pi k a^{-4} \int y^2 dyy f_{\nu_0}(y)\Psi_1 
\label{eq: perturbation_3}
\end{equation}
\begin{equation}
\left(\bar{\rho}_\nu+\bar{P}_\nu\right)\sigma_\nu = \frac{8\pi}{3}a^{-4}\int y^2dy\frac{y^2}{\epsilon_{\nu, \rm eff}}f_0(y)\Psi_2,
\label{eq: perturbation_4}
\end{equation}
where
\begin{equation}
\dot\Psi_0 = -\frac{yk}{\epsilon_{\nu\rm,eff}}\Psi_1-\dot\phi\frac{d\ln f_0}{d\ln y},
\label{eq: Boltzmann_hierarchy1}
\end{equation}
\begin{equation}
\dot\Psi_1 = \frac{yk}{3\epsilon_{\nu\rm,eff}}\left(\Psi_0-2\Psi_2\right) - \frac{\epsilon_{\nu,\rm eff}k}{3y}\psi\frac{d\ln f_0}{d\ln y}
\label{eq: Boltzmann_hierarchy2}
\end{equation}
\begin{equation}
\dot\Psi_l = \frac{yk}{(2l+1)\epsilon_{\nu\rm,eff}}\left[l\Psi_{l-1}-(l+1)\Psi_{l+1}\right].
\label{eq: Boltzmann_hierarchy3}
\end{equation}
For negative effective neutrino masses, the change in the sign of $\epsilon_{\nu,\rm eff}$ also changes the signs of the corresponding terms in the Boltzmann hierarchy equations \eqref{eq: Boltzmann_hierarchy1}--\eqref{eq: Boltzmann_hierarchy3}.
However, these sign changes can be absorbed into the transformation
\[
\widetilde{\Psi}_l=(-1)^l\Psi_l.
\]
Under this transformation, the Boltzmann hierarchy equations in Eqs.~\eqref{eq: Boltzmann_hierarchy1}--\eqref{eq: Boltzmann_hierarchy3} take the same form as those for positive neutrino masses with the same absolute value and can be solved using the same mathematical framework.
Meanwhile, the right hand sides of Eqs.~\eqref{eq: perturbation_1} and ~\eqref{eq: perturbation_2}  acquire an overall minus sign under this transformation.
Consequently, the negative signs cancel in both effective sound speed $c_{s,\nu}^2\equiv\delta P_\nu/\delta\rho_\nu$ and density contrast $\delta_\nu\equiv\delta\rho_\nu/\bar{\rho}_\nu$.
Similarly, the minus signs on the right hand sides of Eqs.~\eqref{eq: perturbation_3} and ~\eqref{eq: perturbation_4} cancel with the negative factor $\left(\bar{\rho}_\nu+\bar{P}_\nu\right)$ on the left hand sides.
Thus, the expressions for $\theta_\nu$ and $\sigma_\nu$ take the same form as those for positive neutrino masses with the same absolute value.
Substituting these relations into Eqs.~\eqref{eq: conti} and \eqref{eq: eular}, it can be found that the evolution equations for $\delta_\nu$ and $\theta_\nu$ remain the same form as those for positive neutrino masses.
It can be therefore expected that $\delta_\nu$s are similar for positive and negative effective neutrino masses with the same absolute value.

To further investigate the free-streaming behavior of effective neutrinos, we turn to the evolution equation for the density contrast $\delta_\nu$ in non-relativistic and sub-horizon limits, neglecting anisotropic stress and the time derivatives of gravitational potentials,
\begin{equation}
\ddot{\delta}_\nu + \frac{\dot a}{a}\dot{\delta}_\nu + k^2c_{s,\nu}^2\delta_\nu\simeq\frac{3}{2}a^2H^2\Omega_m(a)\delta_m,
\label{eq: delta_nu_evol}
\end{equation}
and the free-streaming wavenumber can be approximately expressed as
\begin{equation}
k^2_{\rm fs}\simeq\frac{3}{2}\frac{a^2H^2\Omega_m(a)}{c_{s,\nu}^2}.
\label{eq: freestreamingscale}
\end{equation}
Since changing the sign of the effective neutrino mass does not change the form of $c_{s,\nu}^2$, the free-streaming scales should be similar for positive and negative effective neutrino masses with the same absolute value, assuming the same background expansion histories.
However, the free-streaming effects of positive and negative effective neutrino masses on the $P_{cb}(k)$ are different, as discussed in Sec.~\ref{Effects of Negative Effective Neutrino Masses on Structure Formation and the 21 cm Signal}.

\section{Additional results on parameter degeneracies}
\label{app: degeneracy}

\begin{figure}[htbp]
\centering
\includegraphics[width=0.49\textwidth]{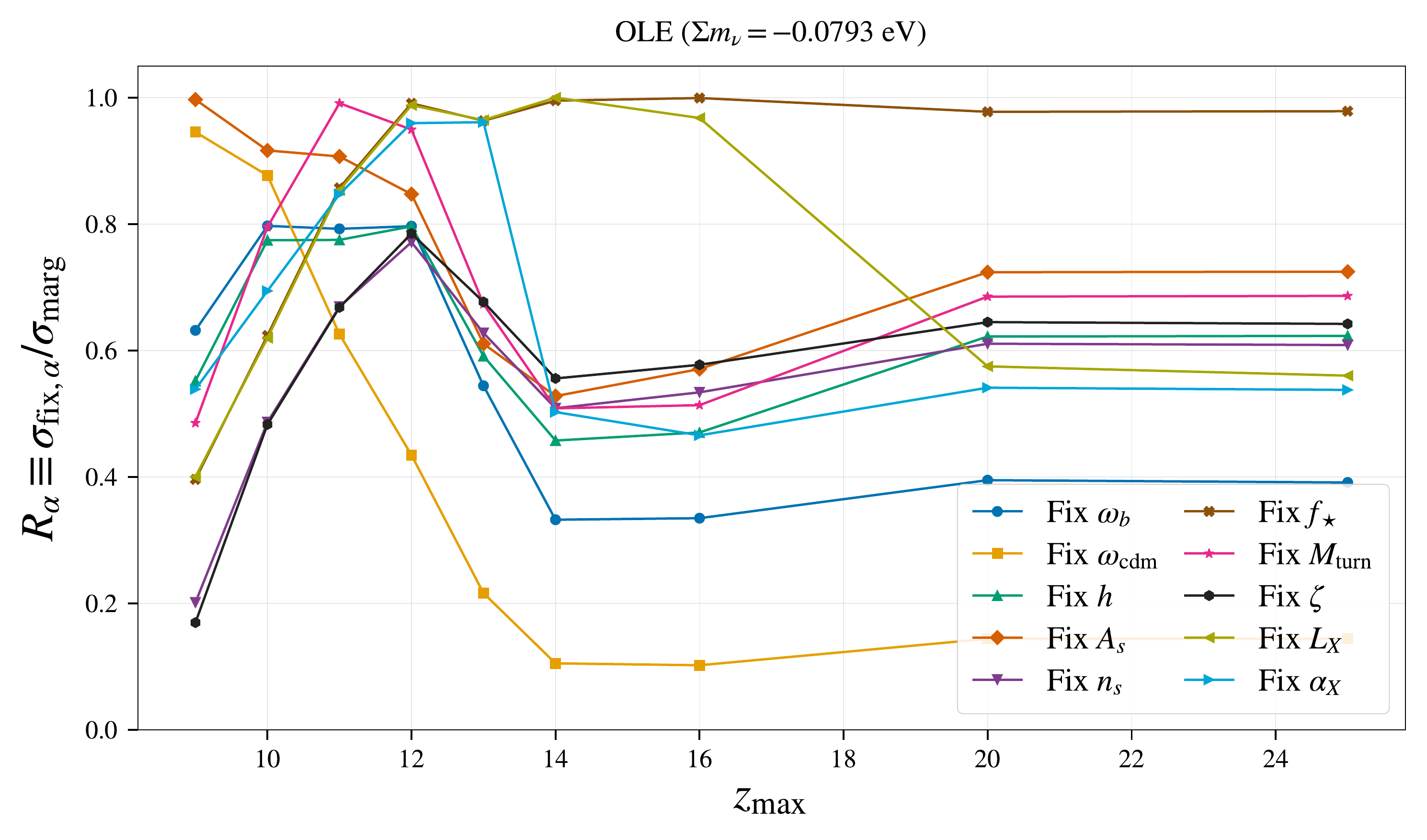}
\includegraphics[width=0.49\textwidth]{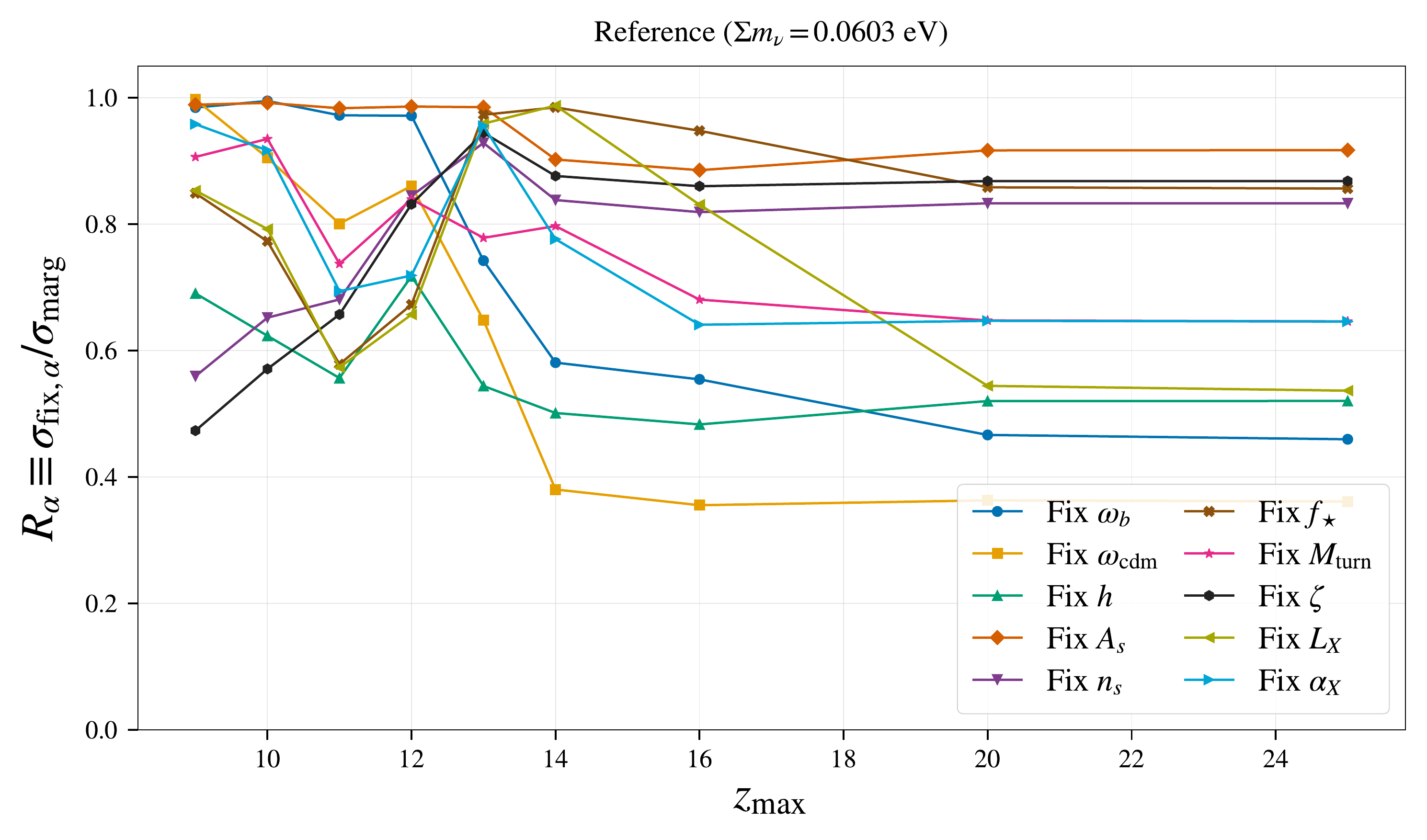}
\includegraphics[width=0.49\textwidth]{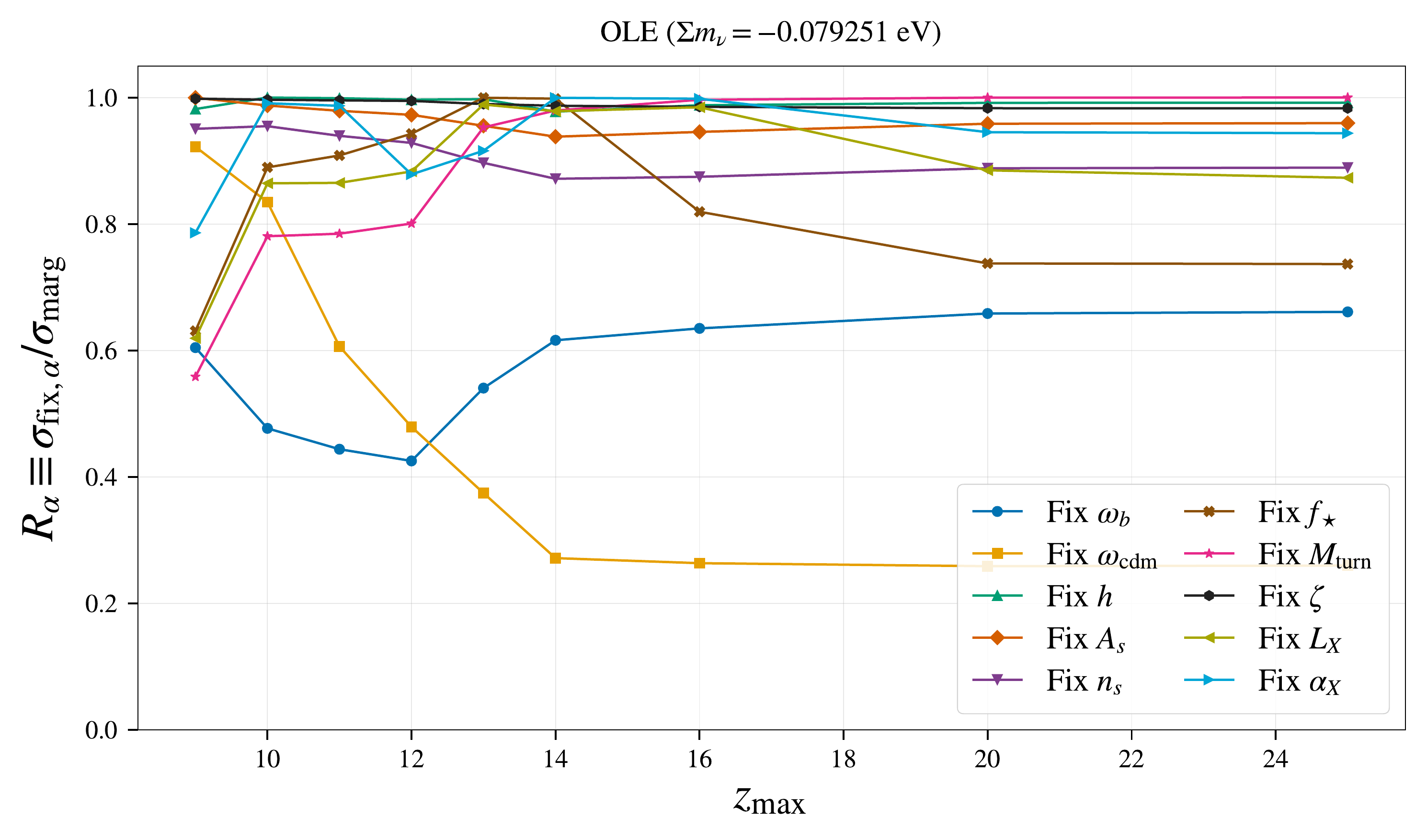}
\includegraphics[width=0.49\textwidth]{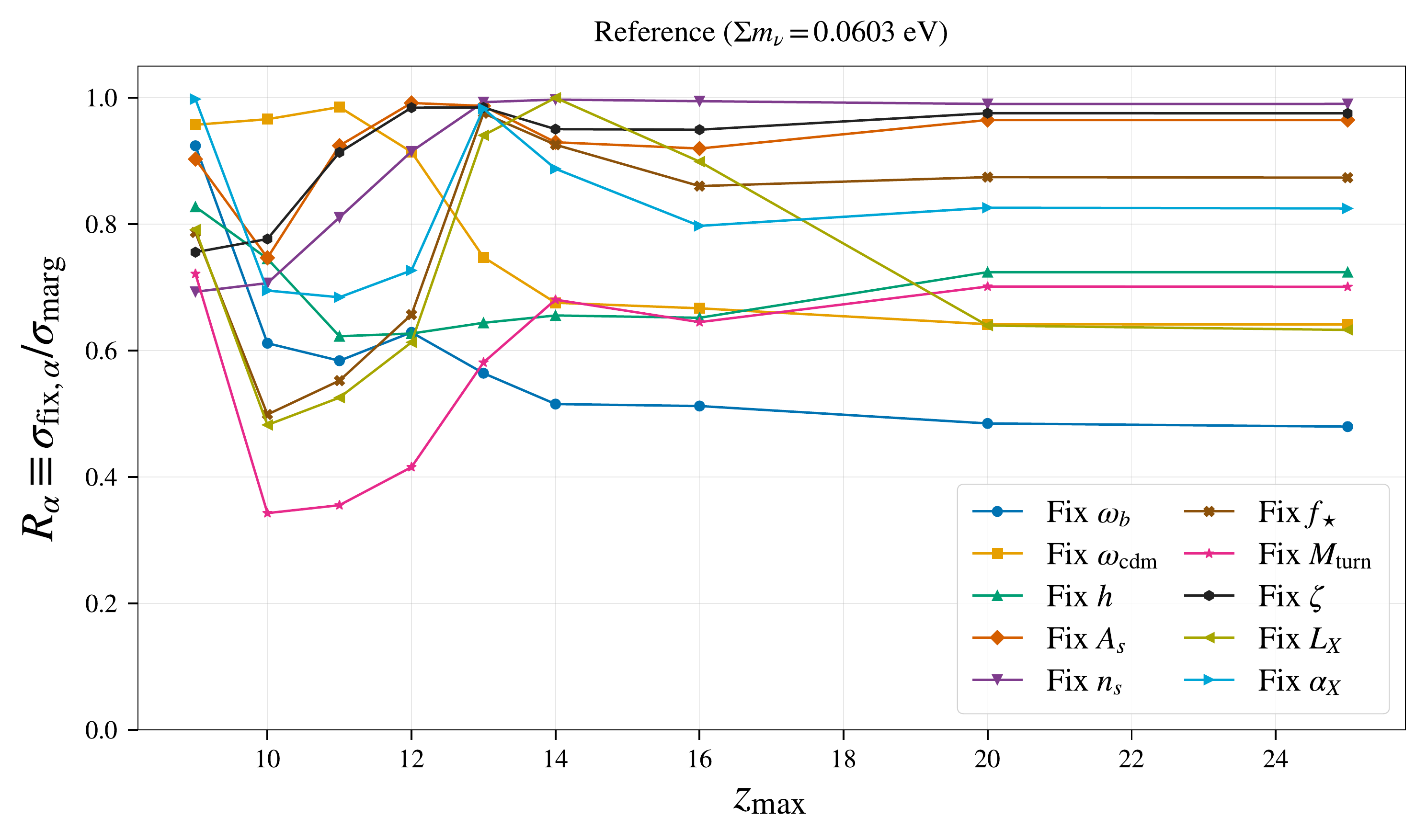}
    \caption{The ratio $R_{\lambda_\alpha}$ defined as a function of the maximum redshift $z_{\max}$ for OLE (left panels) and the Reference cosmology (right panels).
The upper and lower panels correspond to the baseline and $\tau$-matched scenarios, respectively.
Each curve represents the result obtained by fixing one parameter $\lambda_\alpha$ while marginalizing over all remaining parameters.
A smaller $R_{\lambda_\alpha}$ indicates a greater improvement in the neutrino mass constraint. 
}
    \label{fig: degeneracy_R_z}
\end{figure}

\begin{figure}[htbp]
\centering
\includegraphics[width=0.49\textwidth]{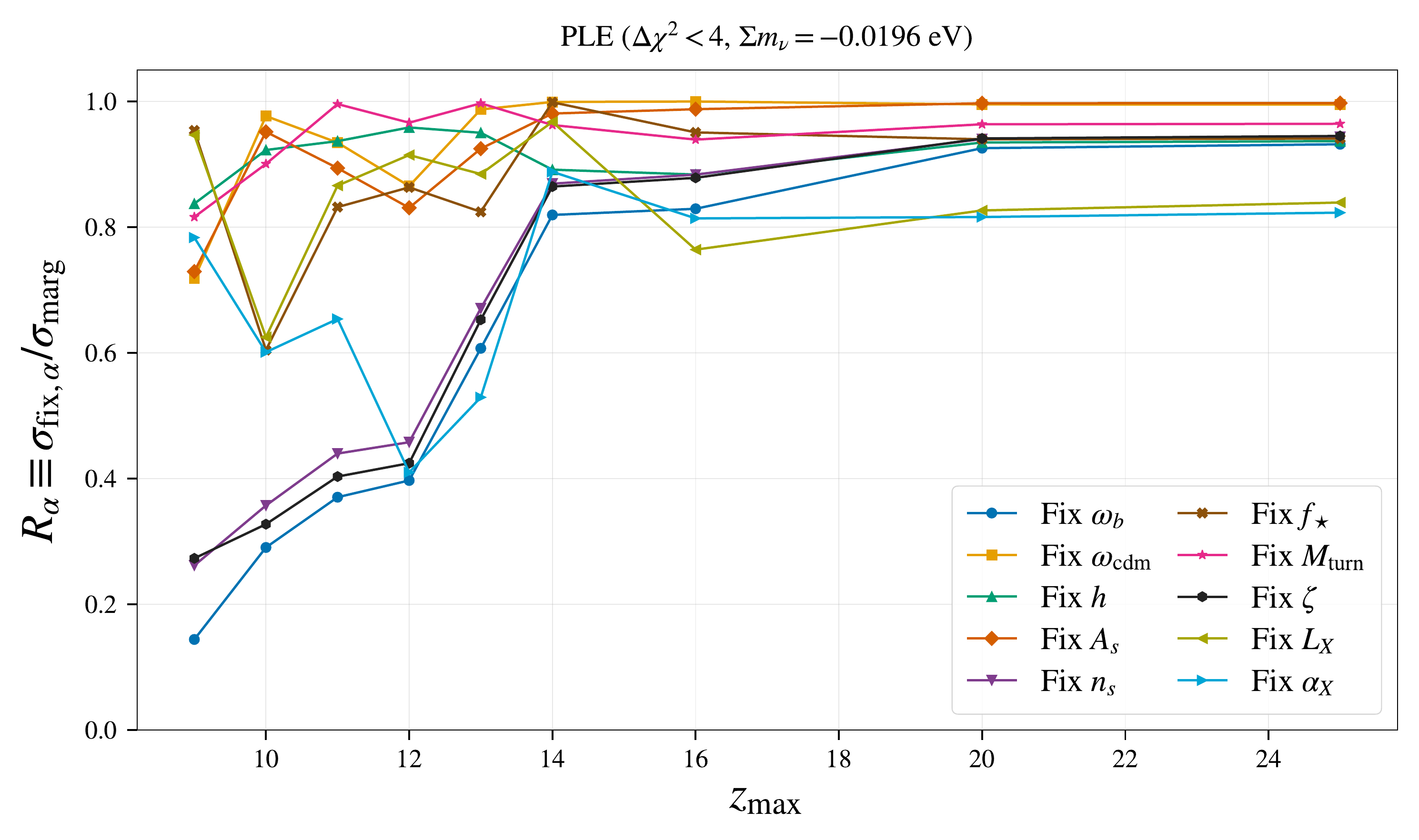}
\includegraphics[width=0.49\textwidth]{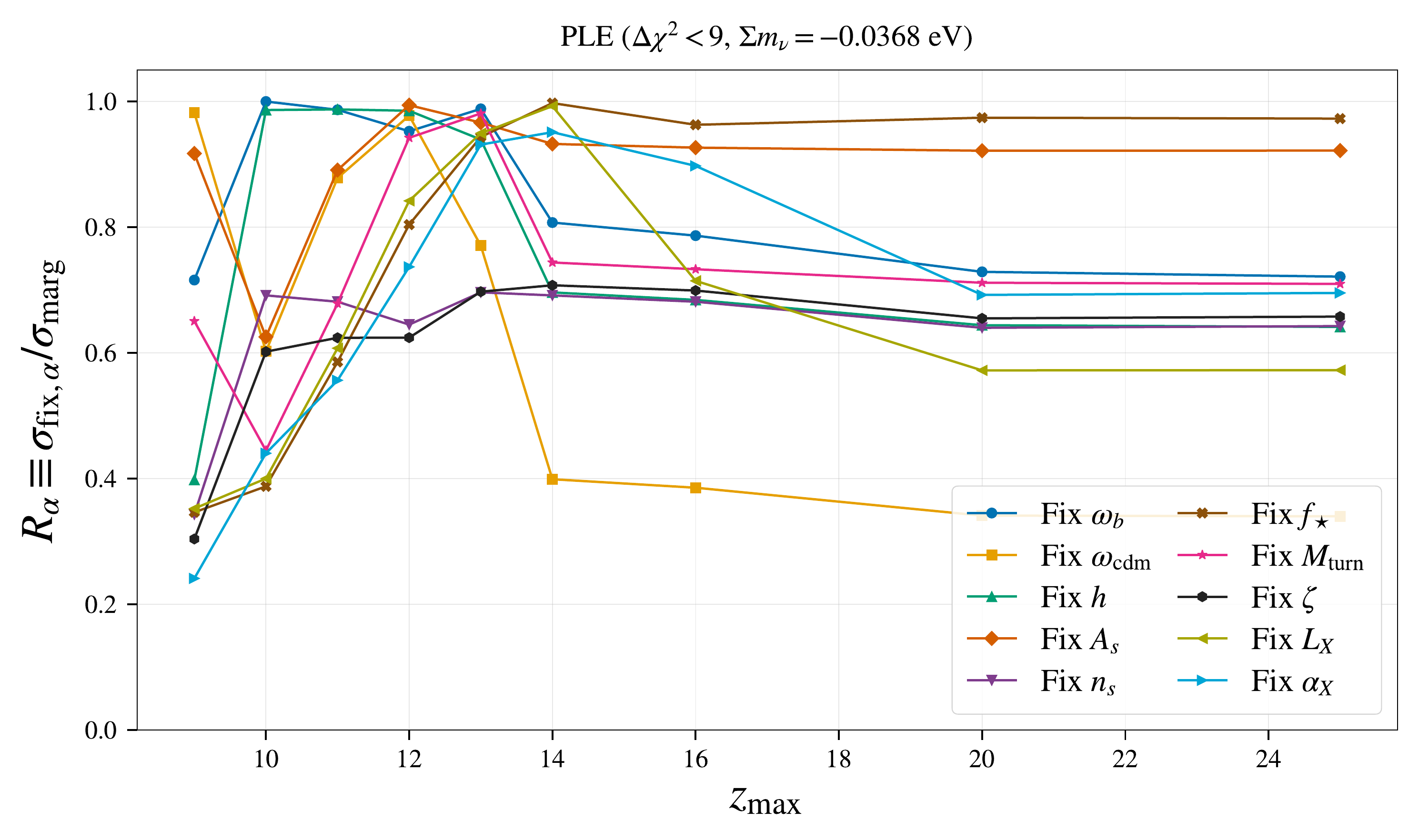}
\includegraphics[width=0.49\textwidth]{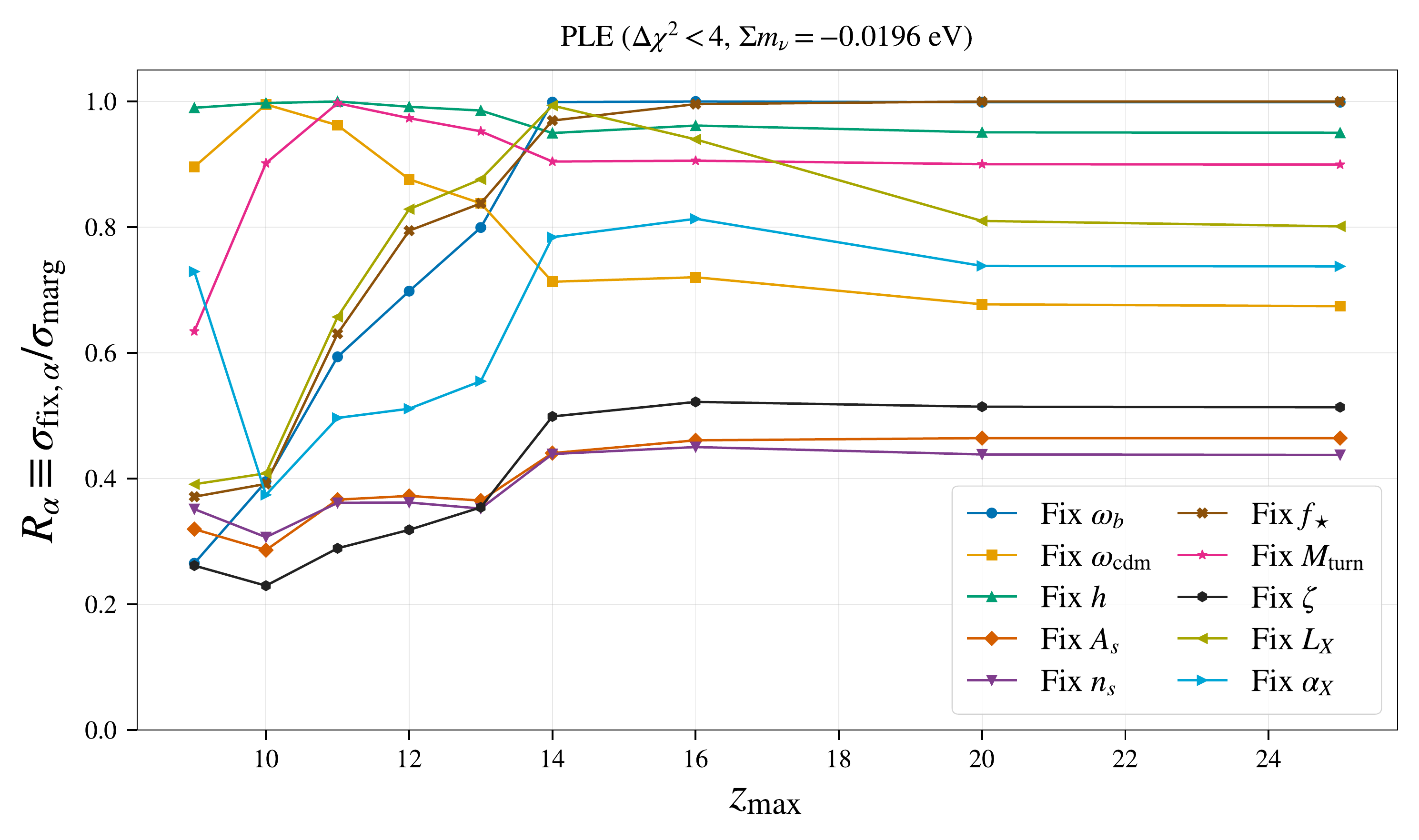}
\includegraphics[width=0.49\textwidth]{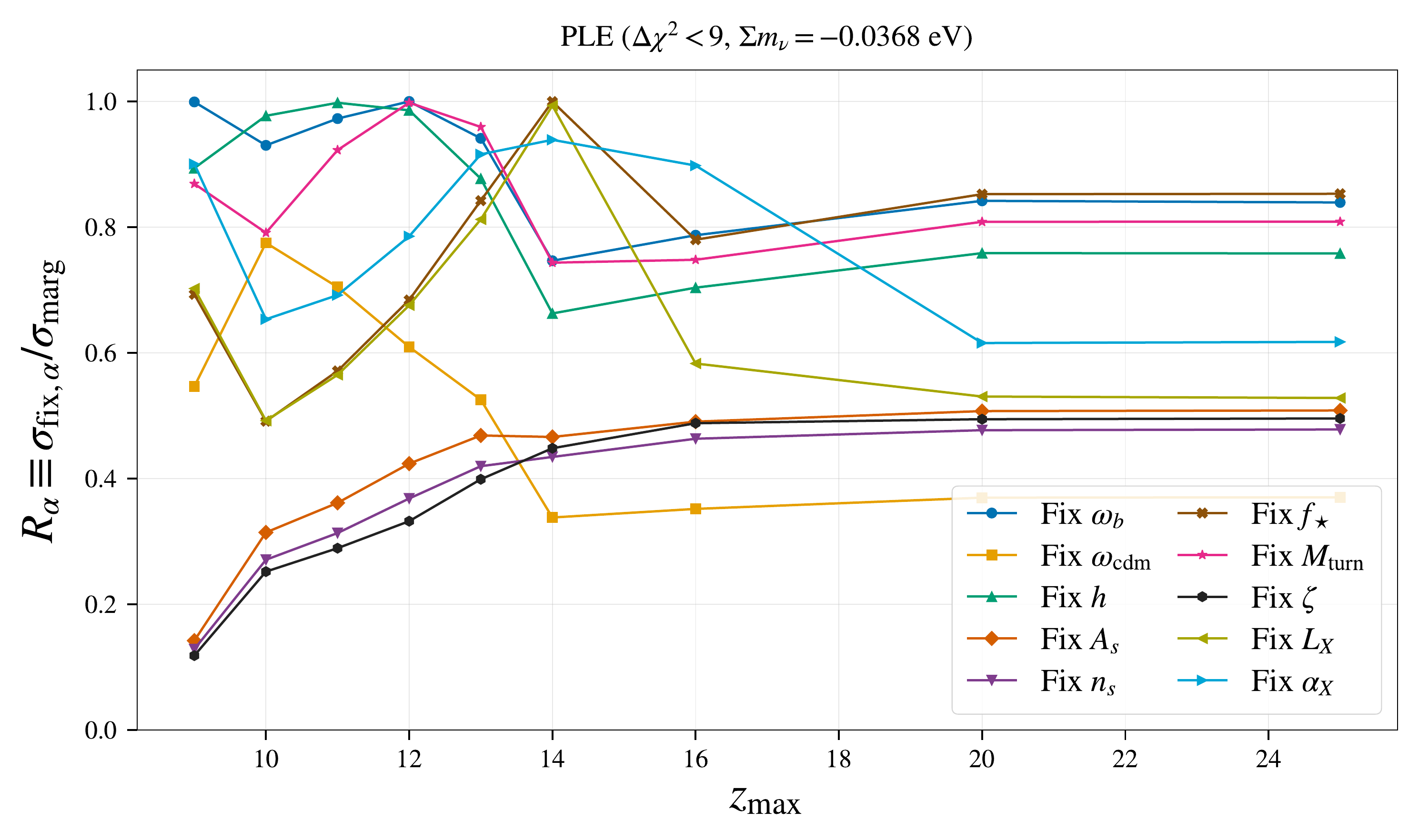}
    \caption{
Same as Fig.~\ref{fig: degeneracy_R_z}, but for the PLE ($\Delta\chi^2<4$) and PLE ($\Delta\chi^2<9$) cosmologies in the left and right panels, respectively.
}
    \label{fig: degeneracy_R_z_app}
\end{figure}

To further investigate the impact of individual parameters on the neutrino mass constraints, we consider fixing one parameter $\lambda_\alpha$ at a time while marginalizing over all the remaining parameters.
The resulting uncertainty for the neutrino mass is denoted by $\sigma_{{\rm fix},\alpha}(\sum m_\nu)$.
We then define the ratio 
\[
R_{\lambda_\alpha} \equiv \frac{\sigma_{{\rm fix},\alpha}(\sum m_\nu)}{\sigma_{{\rm marg}}(\sum m_\nu)},
\]
where a smaller $R_{\lambda_\alpha}$ indicates a greater improvement in the neutrino mass constraint from fixing $\lambda_\alpha$.
The results for OLE and the Reference cosmology under the two scenarios are shown in Fig.~\ref{fig: degeneracy_R_z}, while those for PLE $(\Delta\chi^2<4)$ and PLE $(\Delta\chi^2<9)$ are shown in Fig.~\ref{fig: degeneracy_R_z_app}.
The 21 cm power spectra at different redshifts contain different information about structure formation and astrophysical processes.
Therefore, different parameters may dominate the degeneracies when different redshift ranges are considered.
For example, astrophysical parameters may play a more important role when only low-redshift data are considered, while parameters related to structure formation may become relatively more important as higher redshift data are included.
Among these parameters, $\omega_{\rm cdm}$ has a significant impact on the neutrino mass constraints when higher redshift data are included.
As shown by the orange curves with square markers in the left panels of Fig.~\ref{fig: degeneracy_R_z}, fixing $\omega_{\rm cdm}$ significantly improves the constraints for OLE in both baseline and $\tau$-matched scenarios.
This can be understood because both $\omega_{\rm cdm}$ and $\sum m_\nu$  affect the total matter density and structure formation, allowing their effects to partially compensate each other.
On scales smaller than the neutrino free-streaming scale, positive neutrino masses suppress the growth of matter perturbations.
In contrast, negative effective neutrino masses enhance small-scale structures.
This behavior can be understood from Eqs.~\eqref{eq: delta_m_degenerate} and \eqref{eq: P_cb}.
The details are discussed in Appendix~\ref{Effective Neutrino Free Streaming}.

As shown in Figs.~\ref{fig: degeneracy_R_z} and~\ref{fig: degeneracy_R_z_app}, fixing other cosmological parameters, such as $\omega_b$, $A_s$, and $n_s$, can also reduce the impact of parameter degeneracies and improve the constraints on $\sum m_\nu$.
These results help identify the sources of parameter degeneracies and explain why the Gaussian priors introduced in Sec.~\ref{Impact of External Cosmological Constraints} improve the neutrino mass constraints.
For instance, in the baseline ${\rm PLE}_4$ scenario, as shown in the upper-left panel of Fig.~\ref{fig: degeneracy_R_z_app}, the ratios $R_{\lambda_\alpha}$ remain close to unity at $z_{\max}=25$ when any one of $\omega_b$, $\omega_{\rm cdm}$, $A_s$ and $n_s$ is fixed.
This indicates that more precise constraints on these cosmological parameters have only a limited impact on reducing parameter degeneracies and improving the marginalized constraint on $\sum m_\nu$.
By contrast, in the $\tau$-matched scenario, as shown in the lower-left panel of Fig.~\ref{fig: degeneracy_R_z_app}, $R_{\lambda_\alpha}$ becomes noticeably smaller than unity at $z_{\max}=25$ when $\omega_{\rm cdm}$, $A_s$ or $n_s$ is fixed.
This comparison indicates that more precise constraints on these parameters from independent cosmological probes can more effectively reduce parameter degeneracies and improve the marginalized constraint on $\sum m_\nu$ in the $\tau$-matched scenario.
This helps explain why $Z_{{\rm PLE}_4,{\rm Reference}}$ is larger in the $\tau$-matched scenario than in the baseline scenario, as listed in Tab.~\ref{tab: Z_matrix_4more_accurate}.

We finally turn to the degeneracy involving the astrophysical parameter $\zeta$.
As shown by the black curves in left panels of Fig.~\ref{fig: degeneracy_R_z}, the value of $R_\zeta$ increases from the baseline to the $\tau$-matched scenario.
This indicates that the degeneracy involving $\zeta$ becomes less important in the $\tau$-matched OLE cosmology, helping to explain the improved neutrino mass constraint discussed in Sec.~\ref{Forecast Constraints on Neutrino Mass}.
In contrast, both PLE cases exhibit the opposite behavior as shown in left panel of Fig.~\ref{fig: degeneracy_R_z_app}.
The value of $R_\zeta$ decreases in the $\tau$-matched scenario, indicating that fixing $\zeta$ leads to a greater improvement in the neutrino mass constraints.
Thus, the degeneracy involving $\zeta$ has a stronger impact on the marginalized uncertainties on $\sum m_\nu$ in the $\tau$-matched PLE cases.
These opposite trends help explain why the $\tau$-matched scenario improves the neutrino mass constraint for OLE but weakens those for PLE, as discussed in Sec.~\ref{Forecast Constraints on Neutrino Mass}.

\end{document}